\documentclass[12pt,aps,showpacs,notitlepage]{revtex4-1}
\usepackage{amssymb,amsmath}
\usepackage{bm}
\usepackage[pdftex]{graphicx}
\usepackage{subfig}

\newcommand{\beq}{\begin{equation}}
\newcommand{\beqn}{\begin{eqnarray}}
\newcommand{\eeq}{\end{equation}}
\newcommand{\eeqn}{\end{eqnarray}}
\begin{document}

\title{Primordial Bispectrum from Multifield Inflation with Nonminimal Couplings}
\author{David I. Kaiser, Edward A. Mazenc, and Evangelos I. Sfakianakis}
\email{Email addresses: dikaiser@mit.edu ; mazenc@mit.edu ; esfaki@mit.edu}
\affiliation{Center for Theoretical Physics and Department of Physics, \\
Massachusetts Institute of Technology, Cambridge, Massachusetts 02139 USA}
\date{\today}
\begin{abstract} Realistic models of high-energy physics include multiple scalar fields. Renormalization requires that the fields have nonminimal couplings to the spacetime Ricci curvature scalar, and the couplings can be  large at the energy scales of early-universe inflation. The nonminimal couplings induce a nontrivial field-space manifold in the Einstein frame, and they also yield an effective potential in the Einstein frame with nontrivial curvature. The ridges or bumps in the Einstein-frame potential can lead to primordial non-Gaussianities of observable magnitude. We develop a covariant formalism to study perturbations in such models and calculate the primordial bispectrum. As in previous studies of non-Gaussianities in multifield models, our results for the bispectrum depend sensitively on the fields' initial conditions.
\end{abstract}
\pacs{04.62+v; 98.80.Cq. Published in {\it Physical Review D} 87 (2013): 064004}
\maketitle

\section{Introduction} 

Inflationary cosmology remains the leading account of the very early universe, consistent with high-precision measurements of the cosmic microwave background radiation (CMB) \cite{GuthKaiser,KomatsuWMAP,HinshawWMAP}. A longstanding challenge, however, has been to realize successful early-universe inflation within a well-motivated model from high-energy particle physics. 

Realistic models of high-energy physics routinely include multiple scalar fields \cite{LythRiotto,Mazumdar}. Unlike single-field models, multifield models generically produce entropy (or isocurvature) perturbations. The entropy perturbations, in turn, can cause the gauge-invariant curvature perturbation, $\zeta$, to evolve even on the longest length-scales, after modes have been stretched beyond the Hubble radius during inflation  \cite{preheatingevolution,GWBM,Groot,EastherGiblin,DiMarco,BTW,MalikWands,WandsReview}. Understanding the coupling and evolution of entropy perturbations in multifield models is therefore critical for studying features in the predicted power spectrum, such as non-Gaussianities, that are absent in simple single-field models. (For reviews see \cite{BartoloReview,MalikWands,WandsReview,XingangReview,ByrnesReview,WhitePaper}.)

Recent reviews of primordial non-Gaussianities have emphasized four criteria, at least one of which must be satisfied as a necessary (but not sufficient) condition for observable power spectra to deviate from predictions of single-field models. These criteria include \cite{WhitePaper,XingangReview}: (1) multiple fields; (2) noncanonical kinetic terms; (3) violation of slow-roll; or (4) an initial quantum state for fluctuations different than the usual Bunch-Davies vacuum. As we demonstrate here, the first three of these criteria are {\it generically} satisfied by models that include multiple scalar fields with nonminimal couplings to the spacetime Ricci curvature scalar.

Nonminimal couplings arise in the action as necessary renormalization counterterms for scalar fields in curved spacetime \cite{JackiwColeman,Bunch1980,Fujii,Faraoni2004,BirrellDavies,Buchbinder}. In many models the nonminimal coupling strength, $\xi$, grows without bound under renormalization-group flow \cite{Buchbinder}. In such models, if the nonminimal couplings are $\xi \sim {\cal O} (1)$ at low energies, they will rise to $\xi \gg 1$ at the energy scales of early-universe inflation. We therefore expect realistic models of inflation to incorporate multiple scalar fields, each with a large nonminimal coupling. (Non-Gaussianities in single-field models with nonminimal couplings have been studied in \cite{Qiu}.)

Upon performing a conformal transformation to the Einstein frame --- in which the gravitational portion of the action assumes canonical Einstein-Hilbert form --- the nonminimal couplings induce a field-space manifold that is not conformal to flat \cite{DKconf}. The curvature of the field-space manifold, in turn, can induce additional interactions among the matter fields, beyond those included in the Jordan-frame potential. Moreover, the scalar fields necessarily acquire noncanonical kinetic terms in the Einstein frame. These new features can have a dramatic impact on the behavior of the fields during inflation, and hence on the primordial power spectrum.

Chief among the multifield effects for producing new features in the primordial power spectrum is the ability of fields' trajectories to turn in field-space as the system evolves. Such turns are not possible in single-field models, which include only a single direction of field-space. In the case of multiple fields, special features in the effective potential, such as ridges or bumps, can focus the background fields' trajectories through field space or make them diverge. When neighboring trajectories diverge, primordial bispectra can be amplified to sufficient magnitude that they should be detectable in the CMB \cite{Bernardeau,BartoloReview,Byrnes,ByrnesReview,XingangReview,WandsReview,PTGeometric,SeeryFocusing,Langlois,SeeryLidsey,TanakaMultifield}. 

To date, features like ridges in the effective potential have been studied for the most part phenomenologically rather than being strongly motivated by fundamental physics. Here we demonstrate that ridges arise naturally in the Einstein-frame effective potential for models that incorporate multiple fields with nonminimal couplings. Likewise, as noted above, models with multiple nonminimally coupled scalar fields necessarily include noncanonical kinetic terms in the Einstein frame, stemming from the curvature of the field-space manifold. Both the bumpy features in the potential and the nonzero curvature of the field-space manifold routinely cause the fields' evolution to depart from slow-roll for some duration of their evolution during inflation.

Recent analyses of primordial non-Gaussianities have emphasized two distinct types of fine-tuning needed to produce observable bispectra: fine-tuning the shape of the effective potential to include features like ridges; and separately fine-tuning the fields' initial conditions so that the fields begin at or near the top of these ridges \cite{Byrnes,TanakaMultifield,PTGeometric,SeeryFocusing}. Here we show that the first of these types of fine-tuning is obviated for multifield models with nonminimal couplings; such features of the potential are generic. The second type of fine-tuning, however, is still required: even in the presence of ridges and bumps, the fields' initial conditions must be fine-tuned in order to produce measurable non-Gaussianities.

In Section II we examine the evolution of the fields in the Einstein frame and emphasize the ubiquity of features such as ridges that could make the fields' trajectories diverge in field space. Section III introduces our covariant, multifield formalism for studying the evolution of background fields and linearized perturbations on the curved field-space manifold. In Section IV we analyze adiabiatic and entropy perturbations and quantify their coupling using a covariant version of the familiar transfer-function formalism \cite{BTW,WandsReview,transferfunction}. In Section V we build on recent work \cite{GongTanaka,SeeryBispectrum,ByrnesGong} to calculate the primordial bispectrum for multifield models, applying it here to models with nonminimal couplings. We find that although the nonminimal couplings induce new interactions among the entropy perturbations compared to models in which all fields have minimal coupling, the dominant contribution to the bispectrum remains the familiar local form of $f_{NL}$, made suitably covariant to apply to the curved field-space manifold. Concluding remarks follow in Section VI. We collect quantities relating to the curvature of the field-space manifold in the Appendix.

\section{Evolution in the Einstein Frame}

We consider ${\cal N}$ scalar fields in $(3 + 1)$ spacetime dimensions, with spacetime metric signature $(-, +,+,+)$. We work in terms of the reduced Planck mass, $M_{\rm pl} \equiv (8\pi G)^{-1/2} = 2.43 \times 10^{18}$ GeV. Greek letters label spacetime indices, $\mu, \nu = 0, 1, 2, 3$; lower-case Latin letters label spatial indices, $i, j = 1, 2, 3$; and upper-case Latin letters label field-space indices, $I, J = 1, 2, ...,  {\cal N}$.

In the Jordan frame, the scalar fields' nonminimal couplings to the spacetime Ricci curvature scalar remain explicit in the action. We denote quantities in the Jordan frame with a tilde, such as the spacetime metric, $\tilde{g}_{\mu\nu} (x)$. The action for ${\cal N}$ scalar fields in the Jordan frame may be written
\beq
S_{\rm Jordan} = \int d^4x \sqrt{ - \tilde{g}} \left[ f (\phi^I ) \tilde{R} - \frac{1}{2} \tilde{\cal G}_{IJ} \tilde{g}^{\mu\nu} \partial_\mu \phi^I \partial_\nu \phi^J - \tilde{V} (\phi^I ) \right] ,
\label{SJ}
\eeq
where $f (\phi^I)$ is the nonminimal coupling function and $\tilde{V} (\phi^I)$ is the potential for the scalar fields in the Jordan frame. We have included the possibility that the scalar fields in the Jordan frame have noncanonical kinetic terms, parameterized by coefficients $\tilde{\cal G}_{IJ} (\phi^K)$. Canonical kinetic terms correspond to $\tilde{\cal G}_{IJ} = \delta_{IJ}$. 

We next perform a conformal transformation to work in the Einstein frame, in which the gravitational portion of the action assumes Einstein-Hilbert form. We define a rescaled spacetime metric tensor, $g_{\mu\nu} (x)$, via the relation,
\beq
g_{\mu \nu} (x) = \Omega^2 (x) \> \tilde{g}_{\mu\nu} (x) ,
\label{ghat}
\eeq
where the conformal factor is related to the nonminimal coupling function as
\beq
\Omega^2 (x) = \frac{2}{M_{\rm pl}^2} f (\phi^I (x)) .
\label{Omega}
\eeq
Eq. (\ref{SJ}) then takes the form \cite{DKconf}
\beq
S_{\rm Einstein} = \int d^4 x \sqrt{-g} \left[ \frac{M_{\rm pl}^2}{2} R - \frac{1}{2} {\cal G}_{IJ} g^{\mu\nu} \partial_\mu \phi^I \partial_\nu \phi^J - V (\phi^I) \right] .
\label{SE}
\eeq
The potential in the Einstein frame is scaled by the conformal factor,
\beq
V (\phi^I ) = \frac{1}{\Omega^4 (x)} \tilde{V} (\phi^I ) = \frac{M_{\rm pl}^4}{4f^2 (\phi^I )} \tilde{V} (\phi^I ) .
\label{VE}
\eeq
The coefficients of the noncanonical kinetic terms in the Einstein frame depend on the nonminimal coupling function, $f (\phi^I)$, and its derivatives, and are given by \cite{SalopekBondBardeen,DKconf}
\beq
{\cal G}_{IJ} (\phi^K) = \frac{M_{\rm pl}^2}{2f (\phi^I ) } \left[ \tilde{ \cal G}_{IJ} (\phi^K) + \frac{3}{f (\phi^I )} f_{, I} f_{, J} \right] ,
\label{GIJ}
\eeq
where $f_{, I} = \partial f / \partial \phi^I $.

As demonstrated in \cite{DKconf}, the nonminimal couplings induce a field-space manifold in the Einstein frame, associated with the metric ${\cal G}_{IJ} (\phi^K)$ in Eq. (\ref{GIJ}), which is not conformal to flat for models in which multiple scalar fields have nonminimal couplings in the Jordan frame. Thus there does not exist any combination of conformal transformation plus field-rescalings that can bring the induced metric into the form ${\cal G}_{IJ} = \delta_{IJ}$. In other words, multifield models with nonminimal couplings necessarily include noncanonical kinetic terms in the Einstein frame, even if the fields have canonical kinetic terms in the Jordan frame, $\tilde{\cal G}_{IJ} = \delta_{IJ}$. When analyzing multifield inflation with nonminimal couplings, we therefore must work either with a noncanonical gravitational sector or with noncanonical kinetic terms. Here we adopt the latter. Because there is no way to avoid noncanonical kinetic terms in the Einstein frame in such models, we do not rescale the fields. For the remainder of the paper, we restrict attention to models with canonical kinetic terms in the Jordan frame, $\tilde{\cal G}_{IJ} = \delta_{IJ}$, in which the curvature of the field-space manifold in the Einstein frame depends solely upon $f (\phi^I)$ and its derivatives.

Varying the action of Eq. (\ref{SE}) with respect to $g_{\mu\nu} (x)$ yields the Einstein field equations,
\beq
R_{\mu\nu} - \frac{1}{2} g_{\mu\nu} R = \frac{1}{M_{\rm pl}^2} T_{\mu\nu} ,
\label{EFE1}
\eeq
where 
\beq
T_{\mu\nu} = {\cal G}_{IJ} \partial_\mu \phi^I \partial_\nu \phi^J - g_{\mu\nu} \left[ \frac{1}{2} {\cal G}_{IJ} g^{\alpha \beta} \partial_\alpha \phi^I \partial_\beta \phi^J + V (\phi^I ) \right] .
\label{Tmn}
\eeq
Varying Eq. (\ref{SE}) with respect to $\phi^I$ yields the equation of motion,
\beq
\Box \phi^I + g^{\mu\nu} \Gamma^I_{\>\> JK}  \partial_\mu \phi^J \partial_\nu \phi^K - {\cal G}^{IK} V_{, K} = 0 ,
\label{eomfull}
\eeq
where $\Box \phi^I \equiv g^{\mu\nu} \phi^I_{\>; \mu ; \nu}$ and  $\Gamma^I_{\>\> JK} (\phi^L)$ is the Christoffel symbol for the field-space manifold, calculated in terms of ${\cal G}_{IJ}$. 

We expand each scalar field to first order around its classical background value,
\beq
\phi^I (x^\mu) = \varphi^I (t) + \delta \phi^I (x^\mu) ,
\label{phivarphi}
\eeq
and also expand the scalar degrees of freedom of the spacetime metric to first order, perturbing around a spatially flat Friedmann-Robertson-Walker (FRW) metric \cite{MFB,BTW,MalikWands},
\beq
\begin{split}
ds^2 &= g_{\mu\nu} (x) \> dx^\mu dx^\nu \\
&= - \left(1 + 2A \right) dt^2 + 2a \left( \partial_i B \right) dx^i dt + a^2 \left[ (1 - 2 \psi ) \delta_{ij} + 2 \partial_i \partial_j E \right] dx^i dx^j ,
\end{split}
\label{ds1}
\eeq
where $a(t)$ is the scale factor. To background order, the $00$ and $ij$ components of Eq. (\ref{EFE1}) may be combined to yield the usual dynamical equations,
\beq
\begin{split}
H^2 &= \frac{1}{3 M_{\rm pl}^2} \left[ \frac{1}{2} {\cal G}_{IJ} \dot{\varphi}^I \dot{\varphi}^J + V (\varphi^I)  \right] , \\
\dot{H} &= - \frac{1}{ 2 M_{\rm pl}^2} {\cal G}_{IJ} \dot{\varphi}^I \dot{\varphi}^J ,
\end{split}
\label{Friedmann1}
\eeq
where $H \equiv \dot{a} / a$ is the Hubble parameter, and the field-space metric is evaluated at background order, ${\cal G}_{IJ} = {\cal G}_{IJ } (\varphi^K )$.

Both the curvature of the field-space manifold and the form of the effective potential in the Einstein frame depend upon the nonminimal coupling function, $f (\phi^I)$. The requirement of renormalizability for scalar matter fields in a (classical) curved background spacetime dictates the form of $f (\phi^I)$ \cite{JackiwColeman,Bunch1980,BirrellDavies,Buchbinder}:
\beq
f (\phi^I) = \frac{1}{2} \left[ M_0^2 + \sum_I \xi_I \left( \phi^I \right)^2 \right] ,
\label{f}
\eeq
where $M_0$ is some mass-scale that could be distinct from $M_{\rm pl}$, and the nonminimal couplings $\xi_I$ are dimensionless constants that need not be equal to each other. If any of the fields develop nonzero vacuum expectation values, $\langle \phi^I \rangle = v^I$, then one may expect $M_{\rm pl}^2 = M_0^2 + \sum_I \xi_I ( v^I )^2$. Here we will assume either that $v^I = 0$ for each field or that $\sqrt{ \xi_I}\>  v^I \ll M_{\rm pl}$, so that $M_0 \simeq M_{\rm pl}$. 

The nonminimal couplings $\xi_I$ could in principle take any ``bare" value. (Conformal couplings correspond to $\xi_I = - 1/6$; we only consider positive couplings here, $\xi_I > 0$.) Under renormalization-group flow the constants vary logarithmically with energy scale. The exact form of the $\beta$ functions depends upon details of the matter sector, but for models whose content is akin to the Standard Model the $\beta$ functions are positive and the flow of $\xi_I$ has no fixed point, rising with energy scale without bound \cite{Buchbinder}. Studies of the flow of $\xi$ in the case of Higgs inflation \cite{Higgsinfl} indicate growth of $\xi$ by ${\cal O} (10^1 - 10^2)$ between the electroweak symmetry-break scale, $\Lambda \sim 10^2$ GeV, and typical inflationary scales, $\Lambda \sim 10^{16}$ GeV \cite{Higgsrunning}. Hence we anticipate that realistic models will include nonminimal couplings $\xi_I \gg 1$ during inflation.

Renormalizable potentials in $(3 +1)$ spacetime dimensions can include terms up to quartic powers of the fields. A potential in the Jordan frame that assumes a generic renormalizable, polynomial form such as
\beq
\tilde{V} (\phi^I) = \frac{1}{2} \sum_I m_I^2 \left( \phi^I \right)^2 + \frac{1}{2} \sum_{I < J} g_{IJ} \left( \phi^I \right)^2 \left( \phi^J \right)^2 + \frac{1}{4} \sum_I \lambda_I \left( \phi^I \right)^4 
\label{VtildeIJ}
\eeq
will yield an effective potential in the Einstein frame that is stretched by the conformal factor in accord with Eq. (\ref{VE}). As the $J$th component of $\phi^I$ becomes arbitrarily large the potential in that direction will become asymptotically flat,
\beq
V (\phi^I ) = \frac{M_{\rm pl}^4}{4} \frac{\tilde{V} (\phi^I )}{f^2 (\phi^I )} \rightarrow \frac{M_{\rm pl}^4}{4} \frac{\lambda_J}{ \xi_J^2 } 
\label{VEIJ}
\eeq
(no sum on $J$), unlike the quartic behavior of the potential in the large-field limit in the Jordan frame. (The flatness of the effective potential for large field values was one inspiration for Higgs inflation \cite{Higgsinfl}.) Inflation in such models occurs in a regime of field values such that $\xi_J ( \varphi^J )^2 \gg M_{\rm pl}^2$ for at least one component, $J$. As emphasized in \cite{Higgsinfl}, for large nonminimal couplings, $\xi_J \gg 1$, all of inflation therefore may occur for field values that satisfy $\vert \varphi^J \vert < M_{\rm pl}$, unlike the situation for ordinary chaotic inflation with polynomial potentials and minimal couplings.

Although the effective potential in the Einstein frame will asymptote to a constant value in any given direction of field space, the constants will not, in general, be equal to each other. Thus at finite values of the fields, the potential will generically develop features, such as ridges or bumps, that are absent from the Jordan-frame potential. Because the asymptotic values of $V (\phi^I)$ in any particular direction are proportional to $1/\xi_J^2$, the steepness of the ridges depends sharply on the ratios of the nonminimal coupling constants. If some explicit symmetry, such as the $SU(2)$ electroweak gauge symmetry obeyed by the Higgs multiplet in Higgs inflation \cite{Higgsinfl}, forces all the couplings to be equal --- $\xi_I = \xi$, $m_I^2 = m^2$, and $\lambda_I = g_{IJ} = \lambda$ for all $I, J$ --- then the ridges in the Einstein-frame potential disappear and the potential asymptotes to the same constant value in each direction of field space. We study the dynamics of such special cases in \cite{DKHiggsmultifield}. For the remainder of this paper, we consider models in which the constants are of similar magnitude but not exactly equal to each other. 

\begin{figure}
\centering
\includegraphics[width=4in]{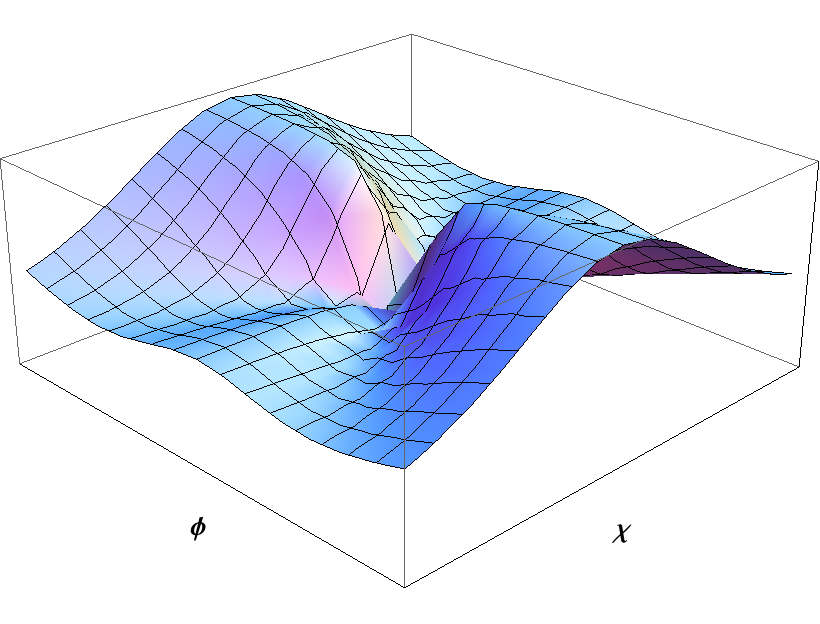}
\caption{\small \baselineskip 14pt The Einstein-frame effective potential, Eq. (\ref{VE2field}), for a two-field model. The potential shown here corresponds to the couplings $\xi_\chi / \xi_\phi = 0.8$,  $\lambda_\chi / \lambda_\phi = 0.3$, $g / \lambda_\phi = 0.1$, and $m_\phi^2 = m_\chi^2 = 10^{-2} \> \lambda_\phi M_{\rm pl}^2$. }
\label{VEinstein}
\end{figure}

For definiteness, consider a two-field model with a potential in the Jordan frame of the form
\beq
\tilde{V} (\phi, \chi) = \frac{1}{2} m_\phi^2 \phi^2 + \frac{1}{2} m_\chi^2 \chi^2  + \frac{1}{2} g \phi^2 \chi^2 + \frac{\lambda_\phi}{4} \phi^4 + \frac{\lambda_\chi}{4} \chi^4 
\label{Vtilde2field}
\eeq
and nonminimal coupling function given by
\beq
f (\phi, \chi) = \frac{1}{2} \left[ M_{\rm pl}^2 + \xi_\phi \phi^2 + \xi_\chi \chi^2 \right].
\label{f2field}
\eeq
In the Einstein frame the potential becomes
\beq
V (\phi, \chi) = \frac{M_{\rm pl}^4}{4} \frac{\left( 2 m_\phi^2 \phi^2 + 2 m_\chi^2 \chi^2 + 2g \phi^2 \chi^2 + \lambda_\phi \phi^4 + \lambda_\chi \chi^4 \right) }{ \left[ M_{\rm pl}^2 + \xi_\phi \phi^2 + \xi_\chi \chi^2 \right]^2} .
\label{VE2field}
\eeq
See Fig. \ref{VEinstein}. 

In addition to the ridges shown in Fig. \ref{VEinstein}, other features of the Einstein-frame potential can arise depending on the Jordan-frame couplings. For example, the tops of the ridges can develop small indentations, such that the top of a ridge along $\chi \sim 0$ becomes a local minimum rather than a local maximum. In that case, field trajectories that begin near the top of a ridge tend to focus rather than diverge, keeping the amplitude of non-Gaussianities very small. For the two-field potential of Eq. (\ref{VE2field}), we find \cite{Footnote} 
\beq
\left( \partial_\chi^2 V \right)_{\vert \chi = 0} = \frac{1}{ \left[ M_{\rm pl}^2 + \xi_\phi \phi^2 \right]^3} \left[ \left( g \xi_\phi - \lambda_\phi \xi_\chi \right) \phi^4 + \left( \xi_\phi m_\chi^2 - 2 \xi_\chi m_\phi^2 + g M_{\rm pl}^2 \right) \phi^2 + m_\chi^2 M_{\rm pl}^2 \right] .
\label{Vchichi}
\eeq
For realistic values of the masses that satisfy $m_\phi^2 , m_\chi^2 \ll M_{\rm pl}^2$, and at early times when $\xi_\phi \phi^2 \gg M_{\rm pl}^2$, the top of the ridge along the $\chi \sim 0$ direction will remain a local maximum if 
\beq
g \xi_\phi < \lambda_\phi \xi_\chi . 
\label{localmax}
\eeq
When the couplings satisfy Eq. (\ref{localmax}), the shape of the potential in the vicinity of its ridges is similar to that of the product potential, $V = m^2 e^{- \lambda \phi^2} \chi^2$, which has been studied in detail in \cite{Byrnes,PTGeometric}. Trajectories of the fields that begin near each other close to the top of a ridge will diverge as the system evolves; that divergence in trajectories can produce a sizeable amplitude for the bispectrum, as we will see below. 

\begin{figure}
\centering
\includegraphics[width=4.5in]{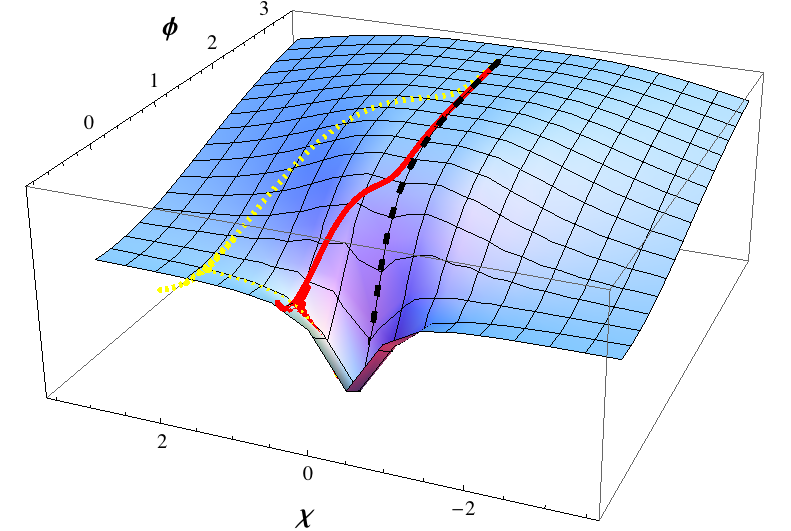}
\caption{\small \baselineskip 14pt Parametric plot of the fields' evolution superimposed on the Einstein-frame potential. Trajectories for the fields $\phi$ and $\chi$ that begin near the top of a ridge will diverge. In this case, the couplings of the potential are $\xi_\phi = 10$, $\xi_\chi = 10.02$, $\lambda_\chi / \lambda_\phi = 0.5$, $g / \lambda_\phi = 1$, and $m_\phi = m_\chi = 0$. (We use a dimensionless time variable, $\tau \equiv \sqrt{\lambda_\phi} \> M_{\rm pl} \> t$, so that the Jordan-frame couplings are measured in units of $\lambda_\phi$.) The trajectories shown here each have the initial condition $\phi (\tau_0) = 3.1$ (in units of $M_{\rm pl}$) and Œdifferent values of $\chi (\tau_0)$: $\chi (\tau_0) = 1.1 \times 10^{-2}$ (``trajectory 1," yellow dotted line); $\chi (\tau_0) = 1.1 \times 10^{-3}$ (``trajectory 2," red solid line); and $\chi (\tau_0) = 1.1 \times 10^{-4}$ (``trajectory 3," black dashed line).}
\label{trajectories123}
\end{figure}

Even potentials with modest ratios of the nonminimal couplings can produce trajectories that diverge sharply, as shown in Fig. \ref{trajectories123}. As we will see in Section V, trajectory 2 of Fig. \ref{trajectories123} (solid red line) yields a sizeable amplitude for the bispectrum that is consistent with present bounds, whereas trajectories 1 and 3 produce negligible non-Gaussianities. We will return to the three trajectories of Fig. 2 throughout the paper, as illustrations of the types of field dynamics that yield interesting possibilities for the power spectrum. 

\begin{figure}
\centering
\includegraphics[width=4.5in]{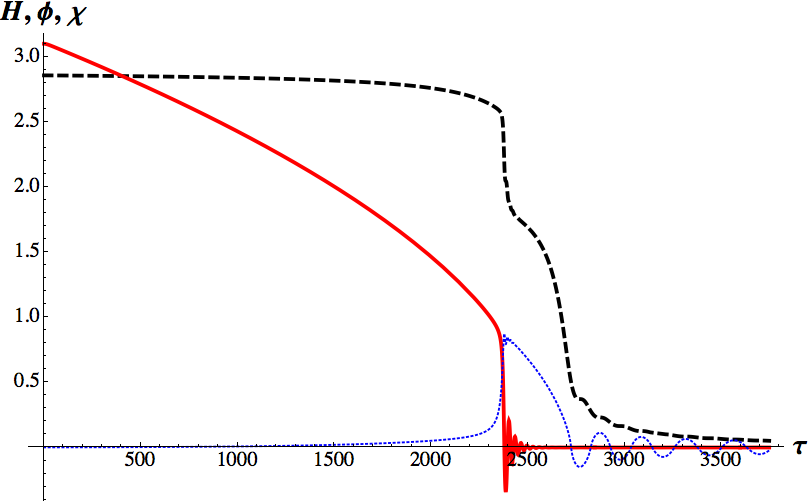}
\caption{\small \baselineskip 14pt The evolution of the Hubble parameter (black dashed line) and the background fields, $\phi (\tau)$ (red solid line) and $\chi (\tau)$ (blue dotted line), for trajectory 2 of Fig. 2. (We use the same units as in Fig. 2, and have plotted $100 H$ so its scale is commensurate with the magnitude of the fields.) For these couplings and initial conditions the fields fall off the ridge in the potential at $\tau = 2373$ or $N = 66.6$ efolds, after which the system inflates for another 4.9 efolds until $\tau_{\rm end} = 2676$, yielding $N_{\rm total} = 71.5$ efolds.}
\label{Hfields}
\end{figure}

Unlike the product potential studied in \cite{Byrnes,PTGeometric}, the potential of Eq. (\ref{VE2field}) contains valleys in which the system will still inflate. For trajectories 1 (orange dotted line) and 2 (red solid line) in Fig. \ref{trajectories123}, for example, the system begins near $\chi \sim 0$ and rolls off the ridge; because $\lambda_\chi / \xi_\chi^2 \neq 0$, the valleys in the $\chi$ direction are also false vacua and hence the system continues to inflate as the fields relax toward the global minimum at $\phi = \chi = 0$. Near the end of inflation, when $\xi_\phi \phi^2 + \xi_\chi \chi^2 < M_{\rm pl}^2$, the fields oscillate around the global minimum of the potential, which can drive a period of preheating. See Fig. \ref{Hfields}. 

Evolution of the fields like that shown in Fig. \ref{Hfields} is generic for this class of models when the fields begin near the top of a ridge, and can produce interesting phenomenological features in addition to observable bispectra. For example, the oscillations of $\phi$ around $\phi = 0$ when the system first rolls off the ridge could produce an observable time-dependence of the scale factor during inflation, as analyzed in \cite{Xingangclock}. The added period of inflation from the false vacuum of the $\chi$ valley could lead to scale-dependent features in the power spectrum associated with double-inflation \cite{doubleinflation}.

\begin{figure}
\centering
\includegraphics[width=4.5in]{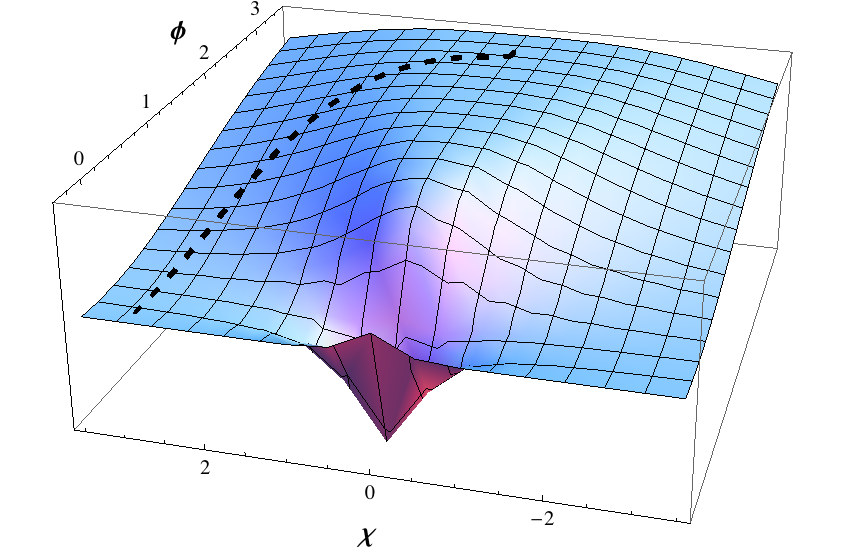}
\caption{ \small \baselineskip 14pt Models with nonzero masses include additional features in the Einstein-frame potential which can also cause neighboring field trajectories to diverge. In this case, we superimpose the evolution of the fields $\phi$ and $\chi$ on the Einstein-frame potential. The parameters shown here are identical to those in Fig. \ref{trajectories123} but with $m^2_\phi = 0.075 \>\lambda_\phi \> M_{\rm pl}^2$ and $m^2_\chi = 0.0025 \> \lambda_\phi \> M_{\rm pl}^2$ rather than 0. The initial conditions match those of trajectory 3 of Fig. \ref{trajectories123}: $\phi (\tau_0) = 3.1$ and $\chi (\tau_0) = 1.1 \times 10^{-4}$ in units of $M_{\rm pl}$.}
\label{trajectorymass}
\end{figure}

In the class of models we consider here, neighboring trajectories may also diverge if we include small but nonzero bare masses for the fields. For example, in Fig. \ref{trajectorymass} we show the evolution of the fields for the same initial conditions as trajectory 3 of Fig. \ref{trajectories123} --- the black, dashed curve that barely deviates from the middle of the ridge. The evolution shown in Fig. \ref{trajectories123} was for the case $m_\phi = m_\chi = 0$. If, instead, we include nonzero masses, then the curvature of the effective potential at small field values becomes different from the zero-mass case. In particular, for positive, real values of the masses, the ridges develop features that push the fields off to one side, recreating behavior akin to what we found in trajectories 1 and 2 of Fig. \ref{trajectories123}. 

\begin{figure}
\centering
\includegraphics[width=4.5in]{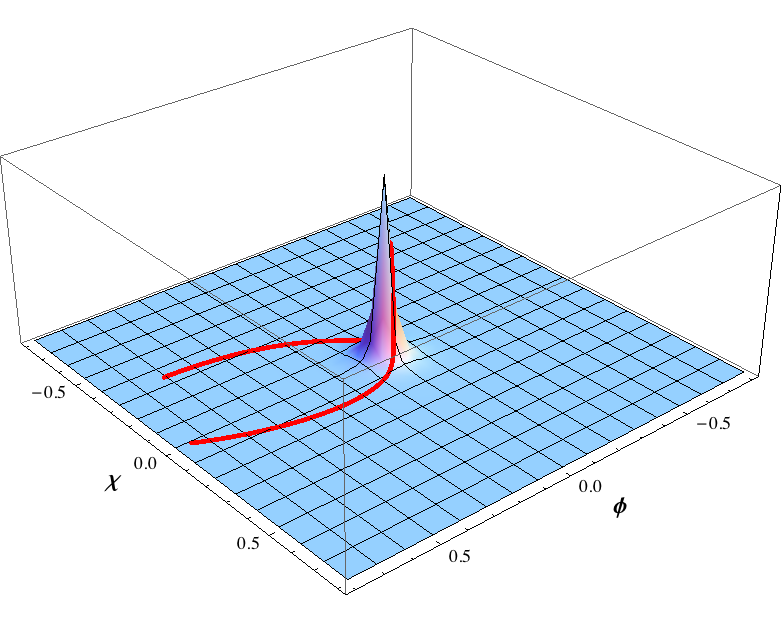}
\caption{\small \baselineskip 14pt Parametric plot of the evolution of the fields $\phi$ and $\chi$ superimposed on the Ricci curvature scalar for the field-space manifold, ${\cal R}$, in the absence of a Jordan-frame potential. The fields' geodesic motion is nontrivial because of the nonvanishing curvature. Shown here is the case $\xi_\phi = 10$, $\xi_\chi = 10.02$, $\phi (\tau_0) = 0.75$, $\chi (\tau_0) = 0.01$, $\phi' (\tau_0) = -0.01$, and $\chi' (\tau_0) = 0.005$.}
\label{Rphichi}
\end{figure}

Because the field-space manifold is curved, the fields' trajectories will turn even in the absence of tree-level couplings from the Jordan-frame potential: the fields' geodesic motion alone is nontrivial. The Ricci scalar for the field-space manifold in the two-field case is given in Eq. (\ref{Ricci2d}). In Fig. \ref{Rphichi} we plot the fields' motion in the curved manifold for the case when $\tilde{V} (\phi, \chi) = V (\phi, \chi) = 0$. The curvature of the manifold is negligible at large field values but grows sharply near $\phi \sim \chi \sim 0$. 

Given the nonvanishing curvature of the field-space manifold, we must study the evolution of the fields and their perturbations with a covariant formalism, to which we now turn.

\section{Covariant Formalism}

A gauge-invariant formalism for studying perturbations in multifield models in the Jordan frame was developed in \cite{DKATT,Sasaki}. In this paper we work in the Einstein frame, following the approach established in \cite{GWBM,Groot,EastherGiblin,DiMarco,Langlois,TanakaMultifield,Byrnes,SeeryFocusing,SeeryLidsey,transferfunction,GongLee,MazumdarWang,PTGeometric,GongTanaka,SeeryBispectrum,ByrnesGong}. Our approach is especially indebted to the geometric formulation of \cite{PTGeometric}. In \cite{PTGeometric}, the authors introduce a particular tetrad construction with which to label the field-space manifold locally, which they dub the ``kinematical basis." The adoption of the kinematical basis simplifies certain expressions and highlights features of physical interest in the primordial power spectrum, but it does so at the expense of obscuring the relationship between observable quantities and the fields that appear in the original Lagrangian, in terms of which any given model is specified. Rather than adopt the kinematical basis here, we develop a covariant approach in terms of a single coordinate chart that covers the entire field manifold. This offers greater insight into the global structure of the manifold, as illustrated in Fig. \ref{Rphichi}. We also keep coordinate labels explicit, which facilitates application of our formalism to the original basis of fields, $\phi^I$, that appears in the governing Lagrangian. Also unlike \cite{PTGeometric}, we work in terms of cosmic time, $t$, rather than the number of efolds during inflation, $N$, because we are interested in applying our formalism (in later work) to eras such as preheating, for which $N$ is a poor dynamical parameter. Because of these formal distinctions from \cite{PTGeometric}, we briefly introduce our general formalism in this section.

We expand each scalar field to first order around its classical background value, as in Eq. (\ref{phivarphi}). The background fields, $\varphi^I (t)$, parameterize classical paths through the curved field-space manifold, and hence can be used as coordinate descriptions of the trajectories. Just like spacetime coordinates in general relativity, $x^\mu$, the array $\varphi^I$ is {\it not} a vector in the field-space manifold \cite{Wald}. Infinitesimal displacements, $d \varphi^I$, do behave as proper vectors, and hence so do derivatives of $\varphi^I$ with respect to an affine parameter such as $t$. 

For any vector in the field space, $A^I$, we define a covariant derivative with respect to the field-space metric as usual by
\beq
{\cal D}_J A^I = \partial_J A^I + \Gamma^I_{\>\> JK} A^K .
\label{covderiv1}
\eeq
Following \cite{Groot,Langlois,PTGeometric}, we also introduce a covariant derivative with respect to cosmic time via the relation
\beq
{\cal D}_t A^I \equiv \dot{\varphi}^J {\cal D}_J A^I = \dot{A}^I + \Gamma^I_{\>\> JK} A^J \dot{\varphi}^K ,
\label{covderivt}
\eeq
where overdots denote derivatives with respect to $t$. The construction of Eq. (\ref{covderivt}) is essentially a directional derivative along the trajectory.

For models with nontrivial field-space manifolds, the tangent space to the manifold at one time will not coincide with the tangent space at some later time. Hence the authors of \cite{GongTanaka,SeeryBispectrum} introduce a covariant means of handling field fluctuations, which we adopt here. As specified in Eq. (\ref{phivarphi}), the value of the physical field at a given location in spacetime, $\phi^I (x^\mu)$, consists of the homogenous background value, $\varphi^I (t)$, and some gauge-dependent fluctuation, $\delta \phi^I (x^\mu)$. The fluctuation $\delta \phi^I$ represents a finite coordinate displacement from the classical trajectory, and hence does not transform covariantly. This motivates a construction of a vector ${\cal Q}^I$ to represent the field fluctuations in a covariant manner. The two field values, $\phi^I$ and $\varphi^I$, may be connected by a geodesic in the field-space manifold parameterized by some parameter $\lambda$, such that $\phi^I (\lambda = 0) = \varphi^I$ and $\phi^I (\lambda = 1) = \varphi^I + \delta \phi^I$. These boundary conditions allow us to identify a unique vector, ${\cal Q}^I$, that connects the two field values, such that ${\cal D}_\lambda \phi^I \vert_{\lambda = 0} = {\cal Q}^I$. One may then expand $\delta \phi^I$ in a power series in ${\cal Q}^I$ \cite{GongTanaka,SeeryBispectrum},
\beq
\delta \phi^I = {\cal Q}^I - \frac{1}{2!} \Gamma^I_{\>\> JK} {\cal Q}^J {\cal Q}^K + \frac{1}{3!} \left( \Gamma^I_{\>\> LM} \Gamma^M_{\>\> JK} - \Gamma^I_{\>\> JK, L} \right) {\cal Q}^J {\cal Q}^K {\cal Q}^L + . . .
\label{deltaphiQ}
\eeq
where the Christoffel symbols are evaluated at background order in the fields, $\Gamma^I_{\>\>JK} = \Gamma^I_{\>\> JK} (\varphi^L)$. To first order in fluctuations $\delta \phi^I \rightarrow {\cal Q}^I$, and hence at linear order we may treat the two quantities interchangeably. When we consider higher-order combinations of the field fluctuations below, however, such as the contribution of the three-point function of field fluctuations to the bispectrum, we must work in terms of the vector ${\cal Q}^I$ rather than $\delta \phi^I$.

We introduce the gauge-invariant Mukhanov-Sasaki variables for the perturbations \cite{MFB,BTW,MalikWands},
\beq
Q^I \equiv {\cal Q}^I + \frac{\dot{\varphi}^I}{H} \psi .
\label{Q}
\eeq
Because both ${\cal Q}^I$ and $\dot{\varphi}^I$ are vectors in the field-space manifold, $Q^I$ is also a vector. The Mukhanov-Sasaki variables, $Q^I$, should not be confused with the vector of field fluctuations, ${\cal Q}^I$. The $Q^I$ are gauge-invariant with respect to spacetime gauge transformations up to first order in the perturbations, and are constructed from a linear combination of field fluctuations and metric perturbations. The quantity ${\cal Q}^I$ does not incorporate metric perturbations; it is constructed from the (gauge-dependent) field fluctuations and background-order quantities such as the field-space Christoffel symbols. At lowest order in perturbations, ${\cal Q}^I \rightarrow Q^I$ in the spatially flat gauge.

Using Eq. ({\ref{Q}), Eq. (\ref{eomfull}) separates into background and first-order expressions,
\beq
{\cal D}_t \dot{\varphi}^I + 3 H \dot{\varphi}^I + {\cal G}^{IK} V_{, K} = 0 ,
\label{eomvarphi}
\eeq
and
\beq
{\cal D}_t^2 Q^I + 3 H {\cal D}_t Q^I + \left[ \frac{k^2}{a^2} \delta^I_{\>\> J} + {\cal M}^I_{\>\> J} - \frac{1}{ M_{\rm pl}^2 a^3} {\cal D}_t \left( \frac{a^3}{H} \dot{\varphi}^I \dot{\varphi}_J \right) \right] Q^J = 0 .
\label{eomQ}
\eeq
The mass-squared matrix appearing in Eq. (\ref{eomQ}) is given by
\beq
{\cal M}^I_{\>\> J} \equiv {\cal G}^{IK} \left( {\cal D}_J {\cal D}_K V \right) - {\cal R}^I_{\>\> LMJ} \dot{\varphi}^L \dot{\varphi}^M ,
\label{MIJ}
\eeq
where ${\cal R}^I_{\>\> LMJ}$ is the Riemann tensor for the field-space manifold. All expressions in Eqs. (\ref{eomvarphi}), (\ref{eomQ}), and (\ref{MIJ}) involving ${\cal G}_{IJ}$, $\Gamma^I_{\>\> JK}$, ${\cal R}^I_{\>\> LMJ}$, and $V$ are evaluated at background order in the fields, $\varphi^I$. 

The system simplifies further if we distinguish between the adiabatic and entropic directions in field space \cite{GWBM}. The length of the velocity vector for the background fields is given by
\beq
\vert \dot{\varphi}^I \vert \equiv \dot{\sigma} = \sqrt{ {\cal G}_{IJ} \dot{\varphi}^I \dot{\varphi}^J } .
\label{dotsigma}
\eeq
Introducing the unit vector,
\beq
\hat{\sigma}^I \equiv \frac{\dot{\varphi}^I}{\dot{\sigma}} ,
\label{hatsigma}
\eeq
the background equations, Eqs. (\ref{Friedmann1}) and (\ref{eomvarphi}), simplify to
\beq
\begin{split}
H^2 &= \frac{1}{3 M_{\rm pl}^2} \left[ \frac{1}{2} \dot{\sigma}^2 + V \right] , \\
\dot{H} &= - \frac{1}{ 2 M_{\rm pl}^2} \dot{\sigma}^2 
\end{split}
\label{Friedmannsigma}
\eeq
and
\beq
\ddot{\sigma} + 3 H \dot{\sigma} + V_{, \sigma} = 0 ,
\label{eomsigma}
\eeq
where we have defined
\beq
V_{, \sigma} \equiv \hat{\sigma}^I V_{, I} .
\label{Vsigma}
\eeq
The background dynamics of Eqs. (\ref{Friedmannsigma}) and (\ref{eomsigma}) take the form of a single-field model with canonical kinetic term, with the exception that $V (\varphi^I)$ in Eqs. (\ref{Friedmannsigma}) and (\ref{eomsigma}) depends on all ${\cal N}$ independent fields, $\varphi^I$.

The directions in field space orthogonal to $\hat{\sigma}^I$ are spanned by
\beq
\hat{s}^{IJ} \equiv {\cal G}^{IJ} - \hat{\sigma}^I \hat{\sigma}^J .
\label{hats}
\eeq
The quantities $\hat{\sigma}^I$ and $\hat{s}^{IJ}$ obey the useful relations
\beq
\begin{split}
\hat{\sigma}_I \hat{\sigma}^I &= 1 , \\
\hat{s}^{IJ} \hat{s}_{IJ} &= {\cal N} - 1 , \\
\hat{s}^I_{\>\> A} \hat{s}^A_{\>\> J} &= \hat{s}^I_{\>\> J} , \\
\hat{\sigma}_I \hat{s}^{IJ} &= 0 \>\> {\rm for \> all} \> J.
\end{split}
\label{orthonorm}
\eeq
Therefore we may use $\hat{\sigma}^I$ and $\hat{s}^{IJ}$ as projection operators to decompose any vector in field space into components along the direction $\hat{\sigma}^I$ and perpendicular to $\hat{\sigma}^I$ as
\beq
A^I = \hat{\sigma}^I \hat{\sigma}_J A^J + \hat{s}^I_{\>\> J} A^J .
\eeq
In particular, $\dot{S}^I \equiv \hat{s}^I_{\>\> J} \dot{\varphi}^J$ vanishes identically, $\dot{S}^I = 0$. Thus all of the dynamics of the background fields are captured by the behavior of $\dot{\sigma}$ and $\hat{\sigma}^I$. 

Given the simple structure of the background evolution, Eqs. (\ref{Friedmannsigma}) and (\ref{eomsigma}), we introduce slow-roll parameters akin to the single-field case. We define
\beq
\epsilon \equiv - \frac{\dot{H}}{H^2} = \frac{3 \dot{\sigma}^2}{ ( \dot{\sigma}^2 + 2 V )}
\label{epsilon}
\eeq
and
\beq
\eta_{\sigma\sigma} \equiv M_{\rm pl}^2 \frac{{\cal M}_{\sigma\sigma}}{V} ,
\label{etasigmasigma}
\eeq
where we have defined
\beq
\begin{split}
{\cal M}_{\sigma J} &\equiv \hat{\sigma}_I  {\cal M}^I_{\>\> J} = \hat{\sigma}^K  \left( {\cal D}_K {\cal D}_J V \right)  , \\
{\cal M}_{\sigma\sigma} &\equiv \hat{\sigma}_I \hat{\sigma}^J {\cal M}^I_{\>\> J} = \hat{\sigma}^K \hat{\sigma}^J \left( {\cal D}_K {\cal D}_J V \right) .
\end{split}
\label{Msigmasigma}
\eeq
The term in ${\cal M}^I_{\>\> J}$ involving ${\cal R}^I_{\>\> LMJ}$ vanishes when contracted with $\hat{\sigma}_I$ or $\hat{\sigma}^J$ due to the first Bianchi identity (since the relevant term is already contracted with $\hat{\sigma}^L \hat{\sigma}^M$), and hence ${\cal M}_{\sigma\sigma}$ is independent of ${\cal R}^I_{\>\> LMJ}$. For trajectory 2 of Fig. \ref{trajectories123} (solid red line), we see that slow-roll is temporarily violated when the fields roll off the ridge of the potential. See Fig. \ref{epsiloneta}.
\begin{figure}
\centering
\includegraphics[width=4.5in]{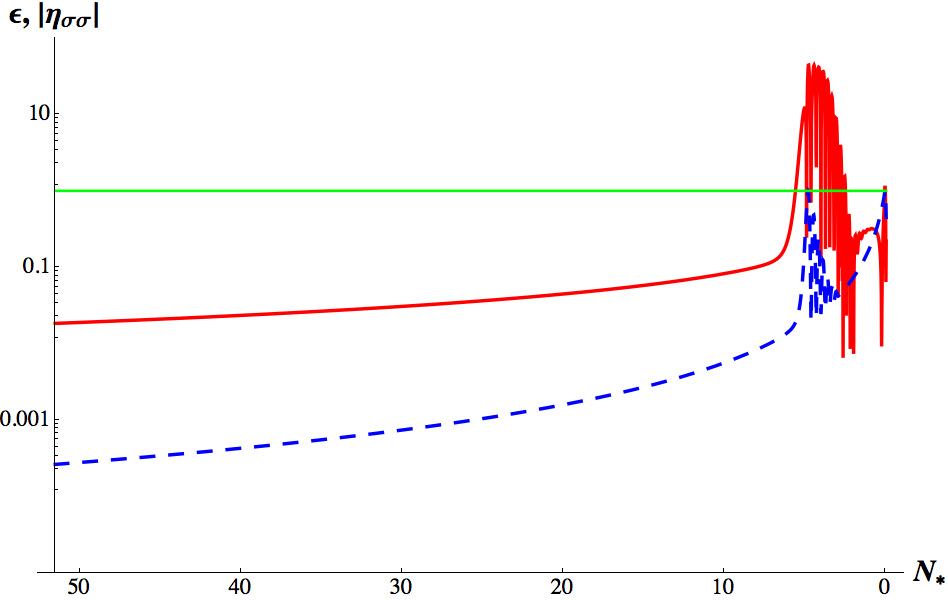}
\caption{\small \baselineskip 14pt The slow-roll parameters $\epsilon$ (blue dashed line) and $\vert \eta_{\sigma\sigma} \vert$ (solid red line) versus $N_*$ for trajectory 2 of Fig. 2, where $N_*$ is the number of efolds before the end of inflation. Note that $\vert \eta_{\sigma\sigma} \vert$ temporarily grows significantly larger than 1 after the fields fall off the ridge in the potential at around $N_* \sim 5$. }
\label{epsiloneta}
\end{figure}

A central quantity of interest is the {\it turn-rate} \cite{PTGeometric}, which we denote $\omega^I$. The turn-rate is given by the (covariant) rate of change of the unit vector, $\hat{\sigma}^I$,
\beq
\omega^I \equiv {\cal D}_t \hat{\sigma}^I = - \frac{1}{\dot{\sigma}} V_{, K} \hat{s}^{IK} ,
\label{omega}
\eeq
where the last expression follows upon using the equations of motion, Eqs. (\ref{eomvarphi}) and (\ref{eomsigma}). Because $\omega^I \propto \hat{s}^{IK}$, we have
\beq
\omega^I \hat{\sigma}_I = 0 .
\eeq
Using Eqs. (\ref{hats}) and (\ref{omega}), we also find
\beq
{\cal D}_t \hat{s}^{IJ} = - \hat{\sigma}^I \omega^J - \omega^I \hat{\sigma}^J .
\eeq
For evolution of the fields like that shown in Fig. \ref{trajectories123}, the turn-rate peaks when the fields roll off the ridge; see Fig. \ref{omegafig}.
\begin{figure}
\centering
\includegraphics[width=4.5in]{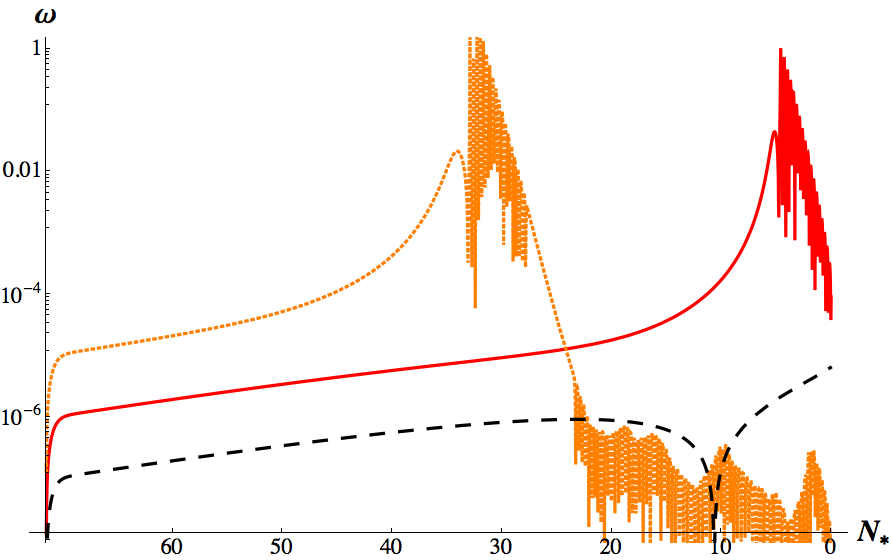}
\caption{\small \baselineskip 14pt The turn-rate, $\omega = \vert \omega^I \vert$, for the three trajectories of Fig. \ref{trajectories123}: trajectory 1 (orange dotted line); trajectory 2 (red solid line); and trajectory 3 (black dashed line). The rapid oscillations in $\omega$ correspond to oscillations of $\phi$ in the lower false vacuum of the $\chi$ valley. For trajectory 1, $\omega$ peaks at $N_* = 34.5$ efolds before the end of inflation; for trajectory 2, $\omega$ peaks at $N_* =  4.9$ efolds before the end of inflation; and for trajectory 3, $\omega$ remains much smaller than 1 for the duration of inflation. }
\label{omegafig}
\end{figure}

We may decompose the perturbations along directions parallel to and perpendicular to $\hat{\sigma}^I$:
\beq
\begin{split}
Q_\sigma &\equiv \hat{\sigma}_I Q^I , \\
\delta s^I &\equiv \hat{s}^I_{\>\> J} Q^J .
\label{QsigmasI}
\end{split}
\eeq
Note that $\delta s^I$ may be defined either in terms of the field fluctuations or the Mukhanov-Sasaki variables, since $\hat{s}^I_{\>\> J} \delta \phi^J = \hat{s}^I_{\>\> J} Q^J$. Though $\delta s^I$ is a vector in field-space with ${\cal N}$ components, only ${\cal N} - 1$ of these components are linearly independent. We will isolate particular components of interest in Section IV.

Taking a Fourier transform, such that for any function $F (t, x^i)$ we have $a^2 (t) \partial_i \partial^i F (t, x^i) = - k^2 F_k(t)$ where $k$ is the comoving wavenumber, Eq. (\ref{eomQ}) separates into two equations of motion (we suppress the label $k$ on Fourier modes),
\beq
\begin{split}
\ddot{Q}_\sigma + 3 H \dot{Q}_\sigma &+ \left[ \frac{k^2}{a^2} + {\cal M}_{\sigma\sigma} - \omega^2 - \frac{1}{M_{\rm pl}^2 a^3} \frac{d}{dt} \left( \frac{a^3 \dot{\sigma}^2}{H} \right) \right] Q_\sigma \\
&\quad\quad\quad = 2 \frac{d}{dt} \left( \omega_J \delta s^J \right) - 2 \left( \frac{ V_{, \sigma}}{\dot{\sigma}} + \frac{\dot{H}}{H} \right) \left( \omega_J \delta s^J \right) 
\end{split}
\label{eomQsigma}
\eeq
and
\beq
\begin{split}
{\cal D}_t^2 \delta s^I + \left[ 3 H \delta^I_{\>\> J} + 2 \hat{\sigma}^I \omega_J \right] {\cal D}_t \delta s^I &+ \left[ \frac{k^2}{a^2} \delta^I_{\>\> J} + {\cal M}^I_{\>\> J} - 2 \hat{\sigma}^I \left( {\cal M}_{\sigma J} + \frac{\ddot{\sigma}}{\dot{\sigma}} \omega_J \right) \right] \delta s^J \\
&\quad\quad\quad = - 2 \omega^I \left[ \dot{Q}_\sigma + \frac{\dot{H}}{H} Q_\sigma - \frac{\ddot{\sigma}}{\dot{\sigma}} Q_\sigma \right] .
\end{split}
\label{eomdeltas1}
\eeq
Although the effective mass of the adiabatic perturbations, $m_{\rm eff}^2 = {\cal M}_{\sigma\sigma} - \omega^2$, is independent of ${\cal R}^I_{\>\> LMJ}$, the curvature of the field-space manifold introduces couplings among components of the entropy perturbations, $\delta s^I$, by means of the ${\cal M}^I_{\>\> J}$ term in Eq. (\ref{eomdeltas1}). The quantities $Q_\sigma$ and $( \omega_J \delta s^J )$ are scalars in field space, so the covariant time derivatives in Eq. (\ref{eomQsigma}) reduce to ordinary time derivatives.

From Eqs. (\ref{eomQsigma}) and (\ref{eomdeltas1}), it is clear that the adiabatic and entropy perturbations decouple if the turn-rate vanishes, $\omega^I = 0$. Moreover, Eq. (\ref{eomQsigma}) for $Q_\sigma$ is identical in form to that of a single-field model (with $m_{\rm eff}^2 = {\cal M}_{\sigma\sigma} - \omega^2$), but with a nonzero source term that depends on the combination $\omega_J \delta s^J$. Even in the presence of significant entropy perturbations, $\delta s^I$, the power spectrum for adiabatic perturbations will be devoid of features such as non-Gaussianities unless the turn-rate is nonzero, $\omega^I \neq 0$.

\section{Adiabatic and Entropy Perturbations}

In Section III we identified the vector of entropy perturbations, $\delta s^I$, which includes ${\cal N} - 1$ physically independent degrees of freedom. As we will see in this section, these ${\cal N} - 1$ physical components may be further clarified by introducing a particular set of unit vectors and projection operators in addition to $\hat{\sigma}^I$ and $\hat{s}^{IJ}$. With them we may identify components of $\delta s^I$ of particular physical interest.

We denote the gauge-invariant curvature perturbation as ${\cal R}_c$, not to be confused with the Ricci scalar for the field-space manifold, ${\cal R}$. The perturbation ${\cal R}_c$ is defined as \cite{MFB,BTW,MalikWands}
\beq
{\cal R}_c \equiv \psi - \frac{H}{(\rho + p )} \delta q ,
\label{Rc1}
\eeq
where $\rho$ and $p$ are the background-order energy density and pressure for the fluid filling the FRW spacetime, and $\delta q$ is the energy-density flux of the perturbed fluid, $T^0_{\>\> i} \equiv \partial_i \delta q$. Given Eq. (\ref{Tmn}), we find 
\beq
\begin{split}
\rho &= \frac{1}{2} \dot{\sigma}^2 + V , \\
p &= \frac{1}{2} \dot{\sigma}^2 - V , \\
\delta q &= - {\cal G}_{IJ} \dot{\varphi}^I \delta \phi^J = - \dot{\sigma} \hat{\sigma}_J \delta \phi^J, 
\end{split}
\label{rhopdeltaq}
\eeq
and hence, upon using Eqs. (\ref{Q}) and (\ref{QsigmasI}),
\beq
{\cal R}_c = \psi + \frac{H}{\dot{\sigma}} \hat{\sigma}_J \delta \phi^J = \frac{H}{\dot{\sigma}} Q_\sigma.
\label{Rc2}
\eeq
We thus find that ${\cal R}_c \propto Q_\sigma$, and that the righthand side of Eq. (\ref{eomdeltas1}) is proportional to $\dot{\cal R}_c$. Recall that these expressions hold to first order in fluctuations, for which $\delta \phi^I \rightarrow {\cal Q}^I$.

In the presence of entropy perturbations, the gauge-invariant curvature perturbation need not remain conserved, $\dot{\cal R}_c \neq 0$. In particular, the nonadiabatic pressure perturbation is given by \cite{BTW,MalikWands}
\beq
\delta p_{\rm nad} \equiv \delta p - \frac{\dot{p}}{\dot{\rho}} \delta \rho = - \frac{2 V_{, \sigma}}{3 H\dot{\sigma}} \delta \rho_m + 2 \dot{\sigma} \left( \omega_J \delta s^J \right) ,
\label{pnad}
\eeq
where $\delta \rho_m \equiv \delta \rho - 3 H \delta q$ is the gauge-invariant comoving density perturbation. The perturbed Einstein field equations (to linear order) require \cite{BTW,MalikWands}
\beq
\delta \rho_m = - 2 M_{\rm pl}^2 \frac{k^2}{a^2} \Psi ,
\label{deltarhom}
\eeq
where $\Psi$ is the gauge-invariant Bardeen potential \cite{MFB,BTW,MalikWands}
\beq
\Psi \equiv \psi + a^2 H \left( \dot{E} - \frac{B}{a} \right) .
\label{Bardeen}
\eeq
Therefore in the long-wavelength limit, for $k \ll aH$, the only source of nonadiabatic pressure comes from the entropy perturbations, $\delta s^I$. Using the usual relations \cite{BTW,MalikWands} among the gauge-invariant quantities ${\cal R}_c$ and $\zeta \equiv -\psi + (H / \dot{\rho}) \delta \rho$, we find
\beq
\dot{\cal R}_c = \frac{H}{\dot{H}} \frac{k^2}{a^2} \Psi + \frac{2H}{\dot{\sigma}} \left( \omega_J \delta s^J \right) .
\label{dotRc}
\eeq
Thus even for modes with $k \ll aH$, ${\cal R}_c$ will not be conserved in the presence of entropy perturbations if the turn-rate is nonzero, $\omega^I \neq 0$.  

Eqs. (\ref{eomQsigma}) and (\ref{dotRc}) indicate that a particular component of the vector $\delta s^I$ is of special physical relevance: the combination $(\omega_J \delta s^J)$ serves as the source for $Q_\sigma$ and hence for $\dot{\cal R}_c$. Akin to the ``kinematical basis" of \cite{PTGeometric}, we may therefore introduce a new unit vector that points in the direction of the turn-rate, $\omega^I$, together with a new projection operator that picks out the subspace perpendicular to both $\hat{\sigma}^I$ and $\omega^I$:
\beq
\begin{split}
\hat{s}^I &\equiv \frac{\omega^I}{\omega} , \\
\gamma^{IJ} &\equiv {\cal G}^{IJ} - \hat{\sigma}^I \hat{\sigma}^J - \hat{s}^I \hat{s}^J ,
\end{split}
\label{sIgammaIJ}
\eeq
where $\omega = \vert \omega^I \vert$ is the magnitude of the turn-rate vector. Using the relations in Eq. (\ref{orthonorm}), the definitions in Eq. (\ref{sIgammaIJ}) imply
\beq
\begin{split}
\hat{s}^{IJ} &= \hat{s}^I \hat{s}^J + \gamma^{IJ} , \\
\gamma^{IJ} \gamma_{IJ} &= {\cal N} - 2 , \\
\hat{s}^{IJ} \hat{s}_J &= \hat{s}^I , \\
\hat{\sigma}_I \hat{s}^I &= \hat{\sigma}_I \gamma^{IJ} = \hat{s}_I \gamma^{IJ} = 0  \>\> {\rm for \> all \>} J .\\
\end{split}
\label{sIorthonorm}
\eeq
We then find
\beq
\begin{split}
{\cal D}_t \hat{s}^I &= - \omega \hat{\sigma}^I - \Pi^I , \\
{\cal D}_t \gamma^{IJ} &= \hat{s}^I \Pi^J + \Pi^I \hat{s}^J
\end{split}
\label{DtsI}
\eeq
where
\beq
\Pi^I \equiv \frac{1}{\omega} {\cal M}_{\sigma K} \gamma^{IK} ,
\label{Pi}
\eeq
and hence, from Eq. (\ref{sIorthonorm}),
\beq
\hat{\sigma}_I \Pi^I = \hat{s}_I \Pi^I = 0 .
\label{Piorthonorm}
\eeq
The vector of entropy perturbations, $\delta s^I$, may then be written as
\beq
\delta s^I =  \hat{s}^I Q_s + B^I , 
\label{QI2}
\eeq
where
\beq
\begin{split}
Q_s &\equiv \hat{s}_J Q^J , \\
B^I &\equiv \gamma^I_{\>\> J} Q^J .
\end{split}
\label{QsBI}
\eeq
The quantity that sources $Q_\sigma$ and ${\cal R}_c$ is now easily identified as the scalar, $\omega_J \delta s^J = \omega Q_s$, which corresponds to just one component of the vector $\delta s^I$.

Making use of Eqs. (\ref{Friedmannsigma}), (\ref{Rc2}), and (\ref{dotRc}), the equation of motion for $\delta s^I$ in Eq. (\ref{eomdeltas1}) separates into
\beq
\begin{split}
\ddot{Q}_s + 3 H \dot{Q}_s &+ \left[ \frac{k^2}{a^2} + {\cal M}_{ss} + 3 \omega^2 - \Pi^2 \right] Q_s \\
&\quad\quad = 4 M_{\rm pl}^2 \frac{\omega}{\dot{\sigma}} \frac{k^2}{a^2} \Psi - {\cal D}_t \left( \Pi_J B^J \right) - \Pi_J {\cal D}_t B^J - {\cal M}_{sJ} B^J - 3 H \left( \Pi_J B^J \right)
\end{split}
\label{eomQs}
\eeq
and
\beq
\begin{split}
{\cal D}_t^2 B^I &+ \left[ 3 H \delta^I_{\>\> J}  + 2 \left( \hat{\sigma}^I \omega_J - \hat{s}^I \Pi_J \right) \right] {\cal D}_t B^J \\
&+ \left[\frac{k^2}{a^2} \delta^I_{\>\> J} + \gamma^{IA} {\cal M}_{AJ} - \hat{\sigma}^I {\cal M}_{\sigma J} - \hat{s}^I \left( 3 H \Pi_J +  {\cal D}_t \Pi_J \right)  \right] B^J \\
&\quad\quad\quad =  2 \Pi^I \dot{Q}_s - \gamma^{IA} {\cal M}_{As} Q_s + \left( 3 H \Pi^I + {\cal D}_t \Pi^I \right) Q_s .
\end{split}
\label{eomBI}
\eeq
In analogy to (\ref{Msigmasigma}), we have introduced the projections
\beq
\begin{split}
{\cal M}_{s J} &\equiv \hat{s}_I {\cal M}^I_{\>\> J} , \\
{\cal M}_{ss} &\equiv \hat{s}_I \hat{s}^J {\cal M}^I_{\>\> J} .
\end{split}
\label{Mss}
\eeq
Note, however, that unlike ${\cal M}_{\sigma J}$, the term in ${\cal M}^I_{\>\> J}$ proportional to ${\cal R}^I_{\>\> LMJ}$ does not vanish upon contracting with $\hat{s}_I$ or $\hat{s}^J$. Hence the Riemann-tensor term in ${\cal M}^I_{\>\> J}$ induces interactions among the components of $\delta s^I$.

For models with ${\cal N} \geq 3$ scalar fields, we may introduce additional unit vectors and projection operators with which to characterize components of $B^I$. The next in the series are given by
\beq
\begin{split}
\hat{u}^I &\equiv \frac{\Pi^I}{\Pi} , \\
q^{IJ} &\equiv \gamma^{IJ} - \hat{u}^I \hat{u}^J .
\end{split}
\label{uIqIJ}
\eeq
Repeating steps as before, we find
\beq
\begin{split}
{\cal D}_t \hat{u}^I &= \Pi \hat{s}^I + \tau^I , \\
{\cal D}_t q^{IJ} &= - \hat{u}^I \tau^J - \tau^I \hat{u}^J ,
\end{split}
\label{DtuI}
\eeq
where
\beq
\tau^I \equiv \frac{1}{ \Pi} \left[ {\cal M}_{s K} + \frac{\dot{\sigma}}{\omega} \hat{\sigma}^A \hat{\sigma}_L \left( {\cal D}_A {\cal M}^L_{\>\> K} \right) \right] q^{IK} .
\label{tauI}
\eeq
We then have
\beq 
B^I = \hat{u}^I Q_u + C^I
\label{BI}
\eeq
in terms of 
\beq
\begin{split}
Q_u &\equiv \hat{u}_J Q^J , \\
C^I &\equiv q^I_{\>\> J} Q^J .
\end{split}
\label{QuCI}
\eeq
This decomposition reproduces the structure in the ``kinematical basis" \cite{PTGeometric} but can be applied in any coordinate basis for the field-space manifold: $Q_\sigma$ is sourced by $Q_s$; $Q_s$ is sourced by $Q_\sigma$ and $Q_u$ (though we have used Eq. (\ref{dotRc}) to substitute the dependence on $\dot{Q}_\sigma$ for the $\nabla^2 \Psi$ term in Eq. (\ref{eomQs})); $Q_u$ is sourced by $Q_s$ and $Q_v \equiv \tau_J Q^J / \vert \tau^I \vert$, and so on.

For our present purposes the two-field model will suffice. The perturbations then consist of two scalar degrees of freedom, $Q_\sigma$ and $Q_s$, which obey Eqs. (\ref{eomQsigma}) and (\ref{eomQs}) (with $B^I = \Pi^I = 0$) respectively. The effective mass-squared of the entropy perturbations becomes
\beq
\mu_s^2 \equiv {\cal M}_{ss} + 3 \omega^2 .
\label{mus}
\eeq
If the entropy perturbations are heavy during slow-roll, with $\mu_s > 3H / 2$, then the amplitude of long-wavelength modes, with $k \ll aH$, will fall exponentially: $Q_s \sim a^{-3/2} (t)$ during quasi-de Sitter expansion. For trajectories that begin near the top of a ridge, on the other hand, the entropy modes will remain light or even tachyonic at early times, since $\mu_s^2$ is related to the curvature of the potential in the direction orthogonal to the background fields' trajectory. Once the background fields roll off the ridge, the entropy mass immediately grows very large, suppressing further growth in the amplitude of $Q_s$. See Fig. \ref{mus2}.
\begin{figure}
\centering
\includegraphics[width=4.5in]{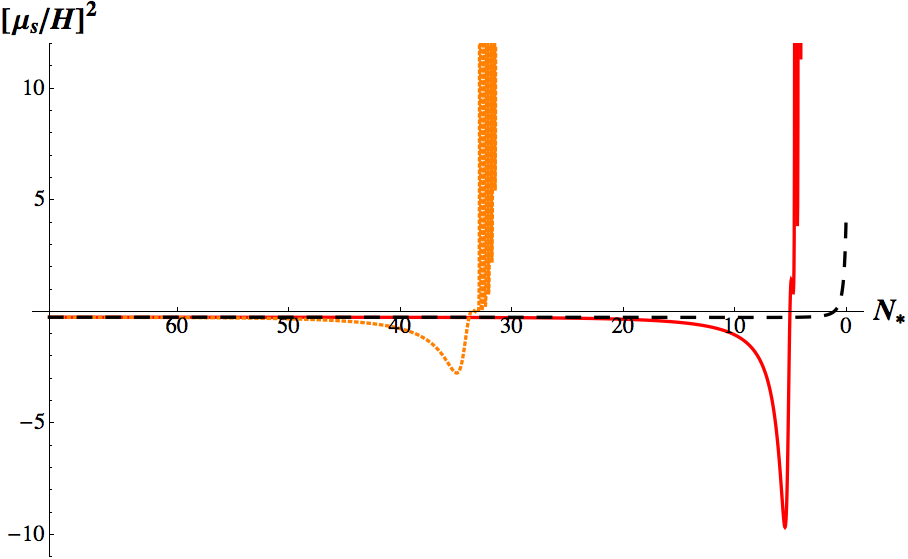}
\caption{\small \baselineskip 14pt The effective mass-squared of the entropy perturbations relative to the Hubble scale, $( \mu_s / H )^2$, for the trajectories shown in Fig. \ref{trajectories123}: trajectory 1 (orange dotted line); trajectory 2 (red solid line); and trajectory 3 (black dashed line). For all three trajectories, $\mu_s^2 < 0$ while the fields remain near the top of the ridge, since $\mu_s^2$ is related to the curvature of the potential in the direction orthogonal to the background fields' evolution. The effective mass grows much larger than $H$ as soon as the fields roll off the ridge of the potential. }
\label{mus2}
\end{figure}

The perturbations in the adiabatic direction are proportional to the gauge-invariant curvature perturbation, as derived in Eq. (\ref{Rc2}). Following the usual convention \cite{transferfunction}, we may define a normalized entropy perturbation as
\beq
{\cal S} \equiv \frac{H}{\dot{\sigma}} Q_s .
\label{calS}
\eeq
In the long-wavelength limit, the coupled perturbations obey general relations of the form \cite{transferfunction}
\beq
\begin{split}
\dot{\cal R}_c &= \alpha H {\cal S} + {\cal O} \left( \frac{k^2}{a^2 H^2} \right) , \\
\dot{\cal S} &= \beta H {\cal S} + {\cal O} \left( \frac{k^2}{a^2H^2} \right) ,
\end{split}
\label{dotRdotS}
\eeq
in terms of which we may write the transfer functions as
\beq
\begin{split}
T_{\cal RS} (t_*, t) &= \int_{t_*}^t dt' \> \alpha (t') H (t') T_{\cal SS} (t_*, t') , \\
T_{\cal SS} (t_*, t) &= \exp \left[ \int_{t_*}^t dt' \> \beta (t') H (t') \right] .
\end{split}
\label{TrsTss}
\eeq
The transfer functions relate the gauge-invariant perturbations at one time, $t_*$, to their values at some later time, $t$. We take $t_*$ to be the time when a fiducial scale of interest first crossed outside the Hubble radius during inflation, defined by $a^2 (t_*) H^2 (t_*) = k_*^2$. In the two-field case, both ${\cal R}_c$ and ${\cal S}$ are scalars in field space, and hence $\alpha$, $\beta$, $T_{\cal RS}$, and $T_{\cal SS}$ are also scalars. Thus there is no time-ordering ambiguity in the integral for $T_{\cal SS}$ in Eq. (\ref{TrsTss}).

In the two-field case, Eq. (\ref{dotRc}) becomes
\beq
\dot{\cal R} = 2 \omega {\cal S} + {\cal O} \left( \frac{k^2}{a^2 H^2} \right) .
\eeq
Comparing with Eq. (\ref{dotRdotS}), we find
\beq
\alpha (t) = \frac{2 \omega (t)}{H (t)} .
\label{alpha}
\eeq
The variation of the gauge-invariant curvature perturbation is proportional to the turn-rate. For $\dot{\cal S}$ we take the long-wavelength and slow-roll limits of Eq. (\ref{eomQs}):
\beq
\dot{Q}_s \simeq - \frac{\mu_s^2}{3H}  Q_s .
\label{dotQs1}
\eeq
Eq. (\ref{dotRdotS}) then yields
\beq
\beta = - \frac{\mu_s^2}{3 H^2}  -  \epsilon + \frac{\ddot{\sigma}}{H \dot{\sigma} }  .
\label{beta1}
\eeq
Taking the slow-roll limit of Eq. (\ref{eomsigma}) for $\dot{\sigma}$, we have
\beq
3 H\dot{\sigma} \simeq - \hat{\sigma}^I V_{, I} .
\label{eomsigmaSR}
\eeq
Taking a covariant time derivative of both sides, using the definition of $\omega^I$ in Eq. (\ref{omega}), and introducing the slow-roll parameter
\beq
\eta_{ss} \equiv M_{\rm pl}^2 \frac{ {\cal M}_{ss} }{V} ,
\label{etass}
\eeq
we arrive at
\beq
\beta = - 2 \epsilon - \eta_{ss} + \eta_{\sigma\sigma} - \frac{4}{3} \frac{\omega^2}{H^2} ,
\label{betafinal}
\eeq
where $\eta_{\sigma\sigma}$ is defined in Eq. (\ref{etasigmasigma}). For trajectories that begin near the top of a ridge, $\eta_{ss}$ will be negative at early times (like $\mu_s^2$), which can yield $\beta > 0$. In that case, $T_{\cal SS} (t_*, t)$ will grow. If one also has a nonzero turn-rate, $\omega$ --- and hence, from Eq. (\ref{alpha}), a nonzero $\alpha$ within the integrand for §$T_{\cal RS} (t_*, t)$ --- then the growing entropy modes will source the adiabatic mode. 

The power spectrum for the gauge-invariant curvature perturbation is defined by \cite{BTW,MalikWands}
\beq
\langle {\cal R}_c ({\bf k}_1) {\cal R}_c ({\bf k}_2 ) \rangle = (2 \pi)^3 \delta^{(3)} ({\bf k}_1 + {\bf k}_2 ) P_{\cal R} (k_1) ,
\label{PRdim}
\eeq
where the angular brackets denote a spatial average and $P_{\cal R} (k) = \vert {\cal R}_c \vert^2$. The dimensionless power spectrum is then given by
\beq
{\cal P}_{\cal R} (k) = \frac{k^3}{2\pi^2} \vert {\cal R}_c \vert^2 ,
\label{PR1}
\eeq
and the spectral index is defined as
\beq
n_s \equiv 1 + \frac{ \partial \ln {\cal P}_{\cal R}}{ \partial \ln k} .
\label{nsdef}
\eeq
Using the transfer functions, we may relate the power spectrum at time $t_*$ to its value at some later time, $t$, as
\beq
{\cal P}_{\cal R} (k) = {\cal P}_{\cal R} (k_*) \left[ 1 + T^2_{\cal RS} (t_*, t)  \right] ,
\label{PR2}
\eeq
where $k$ corresponds to a scale that crossed the Hubble radius at some time $t > t_*$. 
The scale-dependence of the transfer functions becomes \cite{WandsReview,MalikWands,transferfunction,PTGeometric},
\beq
\begin{split}
\frac{1}{H} \frac{\partial T_{\cal RS}}{\partial t_*} &= - \alpha  - \beta  T_{\cal RS} , \\
\frac{1}{H} \frac{\partial T_{\cal SS}}{\partial t_*} &= - \beta  T_{\cal SS} ,
\end{split}
\label{dTRSdt}
\eeq
and hence the spectral index for the power spectrum of the adiabatic fluctuations becomes
\beq
n_s = n_s (t_*) + \frac{1}{H} \left( \frac{\partial T_{\cal RS}}{\partial t_*} \right) \sin \left( 2 \Delta \right) 
\label{ns}
\eeq
where
\beq
\cos \Delta \equiv \frac{T_{\cal RS}}{\sqrt{1 + T_{\cal RS}^2 } } .
\label{Delta}
\eeq
Given Eq. (\ref{eomQsigma}) in the limit $\omega_J \delta s^J = \omega Q_s \ll 1$, the spectral index evaluated at $t_*$ matches the usual single-field result to lowest order in slow-roll parameters \cite{BTW,MalikWands,DKnGETs}:
\beq
n_s (t_*) = 1 - 6 \epsilon (t_*) + 2 \eta_{\sigma\sigma} (t_*) .
\label{nssingle}
\eeq

Scales of cosmological interest first crossed the Hubble radius between 40 and 60 efolds before the end of inflation. In each of the scenarios of Fig. \ref{trajectories123} the fields remained near the top of the ridge in the potential until fewer than 40 efolds before the end of inflation. As indicated in Fig. \ref{Trsfig}, $T_{\cal RS}$ remains small between $N_* = 60$ and $40$ for each of the three trajectories, with little sourcing of the adiabatic perturbations by the entropy perturbations. This behavior of $T_{\cal RS}$ is consistent with the behavior of $\omega = \alpha H / 2$ as shown in Fig. \ref{omegafig}: $\omega$ (and hence $\alpha$) remains small until the fields roll off the ridge in the potential. Only in the case of trajectory 1, which began least high on the ridge among the trajectories and hence fell down the ridge soonest (at $N_* =  34.5$ efolds before the end of inflation), does $T_{\cal RS}$ become appreciable by $N_* = 40$. In particular, we find $T_{\cal RS} (N_{40}) = 0.530$ for trajectory 1; $T_{\cal RS} (N_{40}) = 0.011$ for trajectory 2; and $T_{\cal RS} (N_{40} ) = 0.001$ for trajectory 3. 
\begin{figure}
\centering
\includegraphics[width=4.5in]{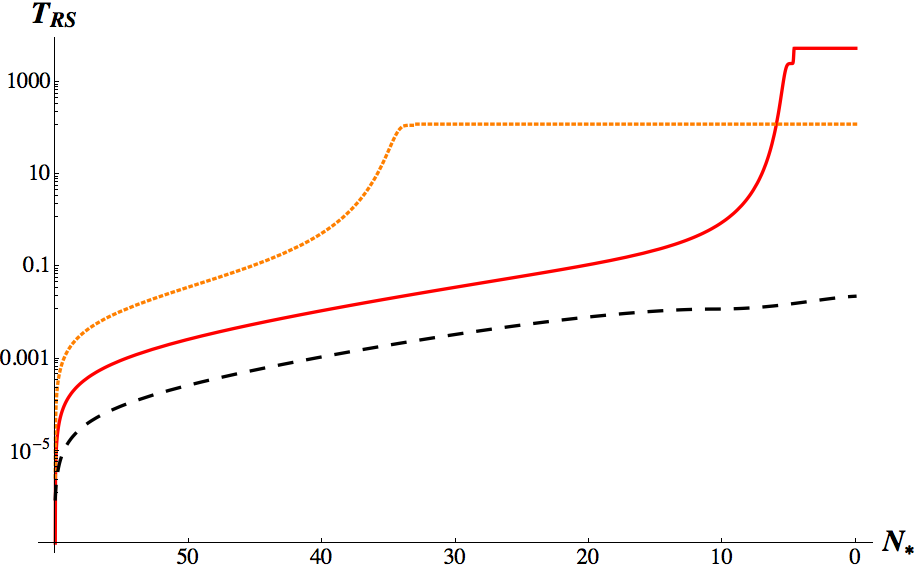}
\caption{\small \baselineskip 14pt The transfer function $T_{\cal RS}$ for the three trajectories of Fig. \ref{trajectories123}: trajectory 1 (orange dotted line); trajectory 2 (red solid line); and trajectory 3 (black dashed line). Trajectories 2 and 3, which begin nearer the top of the ridge in the potential than trajectory 1, evolve as essentially single-field models during early times, before the fields roll off the ridge.}
\label{Trsfig}
\end{figure} 

Fixing the fiducial scale $k_*$ to be that which first crossed the Hubble radius $N_* = 60$ efolds before the end of inflation, we find $n_s (t_*) = 0.967$ for each of the three trajectories of Fig. \ref{trajectories123}, in excellent agreement with the observed value $n_s = 0.971 \pm 0.010$ \cite{HinshawWMAP}. Corrections to $n_s$ from the scale-dependence of $T_{\cal RS}$ remain negligible as long as $T_{\cal RS}$ remains small between $N_* = 60$ and $40$. Consequently, we find negligible tilt in the spectral index across the entire observational window for trajectories 2 and 3, whereas the spectral index for trajectory 1 departs appreciably from $n_s (t_*)$ for scales that crossed the Hubble radius near $N_* = 40$. See Fig. \ref{nsfig}.
\begin{figure}
\centering
\includegraphics[width=4.5in]{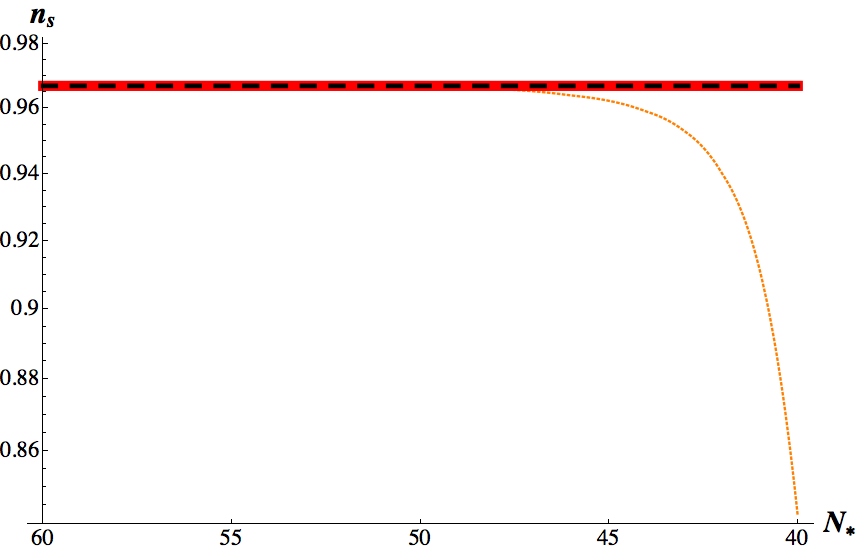}
\caption{\small \baselineskip 14pt The spectral index, $n_s$, versus $N_*$ for the three trajectories of Fig. \ref{trajectories123}: trajectory 1 (orange dotted line); trajectory 2 (red solid line); and trajectory 3 (black dashed line). The spectral indices for trajectories 2 and 3 coincide and show no tilt from the value $n_s (N_{60}) = 0.967$. }
\label{nsfig}
\end{figure}

\section{Primordial Bispectrum}

In the usual calculation of primordial bispectra, one often assumes that the field fluctuations behave as nearly Gaussian around the time $t_*$, in which case the three-point function for the field fluctuations should be negligible. Using the ${\cal Q}^I$ construction of Eq. (\ref{deltaphiQ}), the authors of \cite{GongTanaka,SeeryBispectrum} calculated the action up to third order in perturbations and found several new contributions to the three-point function for field fluctuations, mediated by the Riemann tensor for the field space, ${\cal R}^I_{\>\> JKL}$. The presence of the new terms is not surprising; we have seen that ${\cal R}^I_{\>\> JKL}$ induces new interactions among the perturbations even at linear order, by means of the mass-squared matrix, ${\cal M}^I_{\>\> J}$ in Eq. (\ref{MIJ}). Evaluated at time $t_*$, the three-point function for ${\cal Q}^I$ calculated in \cite{SeeryBispectrum} takes the form
\beq
\begin{split}
\langle {\cal Q}^I ({\bf k}_1 ) {\cal Q}^J ({\bf k}_2 ) {\cal Q}^K ({\bf k}_3 ) \rangle_* &= (2\pi)^3 \delta^{(3)} ( {\bf k}_1 + {\bf k}_2 + {\bf k}_3 ) \frac{H_*^4}{k_1^3 k_2^3 k_3^3 } \\
&\quad\quad \times \left[ {\cal A}_*^{IJK} + {\cal B}_*^{IJK} + {\cal C}_*^{IJK} + {\cal D}_*^{IJK} \right] .
\end{split}
\label{QQQ1}
\eeq
Upon using the definition of $\hat{\sigma}^I$ in Eq. (\ref{hatsigma}), the background equation of Eq. (\ref{Friedmannsigma}) to relate $\dot{H}$ to $\dot{\sigma}^2$, and the definition of $\epsilon$ in Eq. (\ref{epsilon}), the terms on the righthand side of Eq. (\ref{QQQ1}) may be written \cite{SeeryBispectrum,Footnote2}
\beq
\begin{split}
{\cal A}_*^{IJK} &= \frac{\sqrt{2 \epsilon}}{M_{\rm pl} } \> \hat{\sigma}^I {\cal G}^{JK} f_{\cal A} ({\bf k}_1, {\bf k}_2, {\bf k}_3) + {\rm cyclic \> permutations} , \\
{\cal B}_*^{IJK} &= \frac{4 M_{\rm pl} \sqrt{ 2 \epsilon} }{3}  \>  \hat{\sigma}_A {\cal R}^{I (JK) A} f_{\cal B} ({\bf k}_1, {\bf k}_2, {\bf k}_3) + {\rm cyclic \> permutations} , \\
{\cal C}_*^{IJK} &= \frac{2 M_{\rm pl}^2  \> \epsilon}{3} \> \hat{\sigma}_A \hat{\sigma}_B {\cal R}^{(I \vert AB \vert J; K)} f_{\cal C} ({\bf k}_1, {\bf k}_2, {\bf k}_3) + {\rm cyclic \> permutations} , \\
{\cal D}_*^{IJK} &= - \frac{8 M_{\rm pl}^2 \> \epsilon}{3}  \> \hat{\sigma}_A \hat{\sigma}_B {\cal R}^{I (JK) A; B} f_{\cal D} ({\bf k}_1, {\bf k}_2, {\bf k}_3) + {\rm cyclic \> permutations} ,
\end{split}
\label{ABCD}
\eeq
where ${\cal R}^{IABJ;K} = {\cal G}^{KM} {\cal D}_M {\cal R}^{IABJ}$, and $f_{\cal I} ({\bf k}_i )$ are shape-functions in Fourier space that depend on the particular configuration of triangles formed by the wavevectors ${\bf k}_i$. Comparable to the findings in \cite{SeeryLidsey,TanakaMultifield}, each of the contributions to the three-point function for the field fluctuations is suppressed by a power of the slow-roll parameter, $\epsilon \ll 1$.

The quantity of most interest to us is not the three-point function for the field fluctuations but the bispectrum for the gauge-invariant curvature perturbation, $\zeta$, which may be parameterized as
\beq
\langle \zeta ({\bf k}_1 ) \zeta ({\bf k}_2 ) \zeta ({\bf k}_3 ) \rangle \equiv (2 \pi)^3 \delta^{(3)} \left( {\bf k}_1 + {\bf k}_2 + {\bf k}_3 \right) B_\zeta ({\bf k}_1, {\bf k}_2 , {\bf k}_3 ) .
\label{Bzetadef}
\eeq
Recall that the two gauge-invariant curvature perturbations, ${\cal R}_c$ and $\zeta$, coincide in the long-wavelength limit when working to first order in metric perturbations \cite{BTW,MalikWands}. In terms of ${\cal Q}^I$, the $\delta N$ expansion \cite{SasakiStewart,LythMalikSasaki,LythRodriguez,SasakiStewartCovariant} for $\zeta$ on super-Hubble scales becomes \cite{SeeryBispectrum}
\beq
\zeta (x^\mu) = \left( {\cal D}_I  N \right)  {\cal Q}^I (x^\mu) + \frac{1}{2} \left( {\cal D}_I {\cal D}_J N \right) {\cal Q}^I (x^\mu)  {\cal Q}^J (x^\mu) + ...
\label{zetadeltaN}
\eeq
where $N =  \ln \vert a (t_{\rm end}) H (t_{\rm end}) / k_* \vert$ is the number of efolds after a given scale $k_*$ first crossed the Hubble radius until the end of inflation. At $t_*$, Eqs. (\ref{QQQ1}) and (\ref{zetadeltaN}) yield
\beq
\begin{split}
\langle \zeta ({\bf k}_1 ) \zeta ( {\bf k}_2 ) \zeta ( {\bf k}_3 ) \rangle_* &= N_{, I} N_{, J} N_{, K} \langle {\cal Q}^I ({\bf k}_1 ) {\cal Q}^J ( {\bf k}_2  ) {\cal Q}^K ( {\bf k}_3 ) \rangle_* \\
&\quad\quad + \frac{1}{2} \left( {\cal D}_I {\cal D}_J N \right) N_{, K} N_{, L} \\
&\quad\quad\quad \times \int \frac{d^3 q}{(2 \pi)^3 }  \langle {\cal Q}^I ({\bf k}_1 - {\bf q}) {\cal Q}^K ({\bf k}_2 ) \rangle_* \langle {\cal Q}^J ({\bf q}) {\cal Q}^L ({\bf k}_3) \rangle_*  + {\rm cyclic \> perms}  .
\end{split}
\label{zzz1}
\eeq
The bottom two lines on the righthand side give rise to the usual form of $f_{NL}$, made suitably covariant to reflect ${\cal G}_{IJ} \neq \delta_{IJ}$. Adopting the conventional normalization, this term contributes \cite{WandsReview,MalikWands,BartoloReview,XingangReview,ByrnesReview}: 
\beq
\begin{split}
\langle \zeta ({\bf k}_1 ) \zeta ( {\bf k}_2 ) \zeta ( {\bf k}_3 ) \rangle_{f_{NL}} &= (2 \pi)^3 \delta^{(3)} \left({\bf k}_1 + {\bf k}_2 + {\bf k}_3 \right) \frac{H_*^4}{k_1^3 k_2^3 k_3^3} \\
&\quad\quad\quad \times \left[ - \frac{6}{5} f_{NL} \left( N_{,I} N^{,I} \right)^2 \right] \left[ k_1^3 + k_2^3 + k_3^3 \right]
\end{split}
\label{zzzfNL}
\eeq
where
\beq
f_{NL} = - \frac{5}{6} \frac{ N^{, A} N^{, B} {\cal D}_A {\cal D}_B N }{ \left(N_{, I} N^{, I} \right)^2} .
\label{fNL}
\eeq
The term on the first line of Eq. (\ref{zzz1}), proportional to the nonzero three-point function for the field fluctuations, yields new contributions to the bispectrum. However, the three-point function $\langle {\cal Q}^I {\cal Q}^J {\cal Q}^K \rangle_*$ is contracted with the symmetric object, $N_{, I} N_{, J} N_{, K}$. Hence we must consider each term within $A^{IJK}$ with care.

In general, the field-space indices, $I, J, K$, and the momentum-space indices, ${\bf k}_i$, must be permuted as pairs: $(I, {\bf k}_1), (J, {\bf k}_2), (K, {\bf k}_3)$. This is because the combinations arise from contracting the external legs of the various propagators, such as $\langle {\cal Q}^I ({\bf k}_1) {\cal Q}^J ({\bf k}_2) \rangle$ and $\langle {\cal Q}^J ({\bf k}_2) {\cal Q}^K ({\bf k}_3) \rangle$, with the internal legs of each three-point vertex \cite{SeeryBispectrum,Footnote3}. Let us first consider the special case of an equilateral arrangement in momentum space, in which $k_1  = k_2 =  k_3  = k_*$. Then the term proportional to ${\cal A}^{IJK}$ contributes
\beq
\begin{split}
\langle \zeta ({\bf k}_1) \zeta ({\bf k}_2) \zeta ({\bf k}_3) \rangle_{\cal A} &= (2 \pi)^3 \delta^{(3)} ( {\bf k}_1 + {\bf k}_2 + {\bf k}_3 ) \frac{H_*^4}{4 k_*^9} \\
&\quad\quad\quad \times \frac{\sqrt{2 \epsilon}}{M_{\rm pl}} \left( N_{, I} N_{, J} N_{, K} \right)  \left[ \hat{\sigma}^I {\cal G}^{JK} + \hat{\sigma}^J {\cal G}^{KI} + \hat{\sigma}^K {\cal G}^{IJ} \right] f_{\cal A} (k) \\
&= (2 \pi)^3 \delta^{(3)} ( {\bf k}_1 + {\bf k}_2 + {\bf k}_3 ) \frac{H_*^4}{4 k_*^9} \\
&\quad\quad\quad  \times  \frac{3\sqrt{2 \epsilon}}{M_{\rm pl}}  \left[  \left( \hat{\sigma}^I N_{, I} \right) \left( N_{, A} N^{, A} \right) \right] f_{\cal A} (k)  ,
\end{split}
\label{zzzA}
\eeq
where $f_{\cal A} (k)$ depends only on $k$. Taking the equilateral limit of the relevant expression in Eq. (3.17) of \cite{SeeryBispectrum}, we find $f_{\cal A} (k) \rightarrow -5 k_*^3 / 4$. Using Eqs. (\ref{covderivt}), (\ref{hatsigma}),  (\ref{Friedmannsigma}), and (\ref{epsilon}), we also have
\beq
\hat{\sigma}^I N_{, I} = \frac{1}{\dot{\sigma} } \dot{\varphi}^I {\cal D}_I N =\frac{1}{ \dot{\sigma} }{\cal D}_t N = \frac{H}{\dot{\sigma}}  = \frac{1}{M_{\rm pl} H \sqrt{2\epsilon} } ,
\label{hatsigmaN}
\eeq
and hence
\beq
\langle \zeta ({\bf k}_1) \zeta ({\bf k}_2) \zeta ({\bf k}_3) \rangle_{\cal A} = (2 \pi)^3 \delta^{(3)} ( {\bf k}_1 + {\bf k}_2 + {\bf k}_3 ) \frac{H_*^4}{4 k_*^9} \left[ \frac{3}{M_{\rm pl}^2 } \left( N_{, A} N^{,A} \right) \right] f_{\cal A} (k) .
\label{zzzAfinal}
\eeq

The term arising from ${\cal B}^{IJK}$ contributes
\beq
\begin{split}
\langle \zeta ({\bf k}_1) \zeta ({\bf k}_2) \zeta ({\bf k}_3) \rangle_{\cal B} &= (2 \pi)^3 \delta^{(3)} ( {\bf k}_1 + {\bf k}_2 + {\bf k}_3 ) \frac{H_*^4}{4 k_*^9} \\
&\quad\quad \times \frac{4 M_{\rm pl} \sqrt{2 \epsilon} }{3}  \> \hat{\sigma}_A N_{, I} N_{, J} N_{, K} \left[  {\cal R}^{IJKA} + {\cal R}^{IKJA}  + {\rm cyclic} \right] f_{\cal B} (k) .
\end{split}
\label{zzzB1}
\eeq
But from the symmetry properties of the Riemann tensor we have ${\cal R}^{IJKA} = {\cal R}^{KAIJ} = - {\cal R}^{AKIJ}$, and from the first Bianchi identity,
\beq
{\cal R}_{A [ KIJ ] } = 0 .
\label{Bianchi1}
\eeq
The antisymmetry of the Riemann tensor in its last three indices means that any contraction of the form
\beq
{\cal O}_{IJK} {\cal R}^{A KIJ  } = 0
\label{Bianchivanish1}
\eeq
for objects ${\cal O}_{IJK}$ that are symmetric in the indices $I, J, K$. In our case, we have ${\cal O}_{IJK} = N_{, I} N_{, J } N_{, K}$ and thus every term in the square brackets of Eq. (\ref{zzzB1}) including the cyclic permutations may be put in the form of Eq. (\ref{Bianchivanish1}). We therefore find
\beq
\langle \zeta ({\bf k}_1) \zeta ({\bf k}_2) \zeta ({\bf k}_3) \rangle_{\cal B} = 0
\label{zzzBfinal}
\eeq
identically in the equilateral limit.

The term arising from ${\cal C}^{IJK}$ contributes
\beq
\begin{split}
\langle \zeta ({\bf k}_1) \zeta ({\bf k}_2) \zeta ({\bf k}_3) \rangle_{\cal C} &= (2 \pi)^3 \delta^{(3)} ( {\bf k}_1 + {\bf k}_2 + {\bf k}_3 ) \frac{H^4}{4 k_*^9} \\
&\quad\quad \times \frac{2 M_{\rm pl}^2 \> \epsilon}{3}  \> \hat{\sigma}_A \hat{\sigma}_B N_{, I} N_{, J} N_{, K} {\cal R}^{(I \vert AB \vert J ; K ) } f_{\cal C} (k)  .
\end{split}
\label{zzzC1}
\eeq
In the equilateral limit, we find $f_{\cal C} (k_*) \simeq 15 k_*^3$, based on the limit of the appropriate expression in Eq. (3.17) of \cite{SeeryBispectrum}. We may identify the nonzero terms in Eq. (\ref{zzzC1}) using the Bianchi identities. The first Bianchi identity is given in Eq. (\ref{Bianchi1}), and the second Bianchi identity may be written
\beq
{\cal R}_{AB [ CD ; E] } = 0 .
\label{Bianchi2}
\eeq
Using the (anti)symmetry properties of the Riemann tensor and Eqs. (\ref{Bianchi1}) and (\ref{Bianchi2}), together with the fact that the combinations ${\cal O}_{IJK} \equiv N_{, I} N_{, J} N_{, K}$ and $\Omega_{AB} \equiv \hat{\sigma}_A \hat{\sigma}_B$ are symmetric in their indices, we find the only nonzero term within Eq. (\ref{zzzC1}) to be
\beq
\begin{split}
\langle \zeta ({\bf k}_1) \zeta ({\bf k}_2) \zeta ({\bf k}_3) \rangle_{\cal C} &=  (2 \pi)^3 \delta^{(3)} ( {\bf k}_1 + {\bf k}_2 + {\bf k}_3 ) \frac{H^4}{4 k_*^9} \\
&\quad\quad \times \frac{2 M_{\rm pl}^2 \> \epsilon }{3} \> \hat{\sigma}_A \hat{\sigma}_B N_{, I} N_{, J} N_{, K} {\cal R}^{IABJ; K} f_{\cal C} (k) .
\end{split}
\label{zzzCfinal}
\eeq

The final term to consider arises from ${\cal D}^{IJK}$. In particular, in the equilaterial limit we have
\beq
\begin{split}
\langle \zeta ({\bf k}_1) \zeta ({\bf k}_2) \zeta ({\bf k}_3) \rangle_{\cal D} &= (2 \pi)^3 \delta^{(3)} ( {\bf k}_1 + {\bf k}_2 + {\bf k}_3 ) \frac{H^4}{4 k_*^9} \\
&\quad\quad \times
 - \frac{4 M_{\rm pl}^2 \> \epsilon}{3}  \>  \hat{\sigma}_A \hat{\sigma}_B N_{, I} N_{, J} N_{, K} \left[  {\cal R}^{IJKA; B} + {\cal R}^{IKJA; B} + {\rm cyclic} \right]  f_{\cal D} (k) .
\end{split}
\label{zzzD1}
\eeq
Again we may use ${\cal R}_{IJKA} = {\cal R}_{KAIJ} = - {\cal R}_{AKIJ}$ and Eq. (\ref{Bianchi1}) to put the first term in square brackets in Eq. (\ref{zzzD1}) in the form
\beq
{\cal O}_{IJK} {\cal R}^{A KIJ; B} = 0
\eeq
for ${\cal O}_{IJK}$ symmetric. The same occurs for the second term in square brackets in Eq. (\ref{zzzD1}) and for all cyclic permutations of $I, J, K$. Hence we find
\beq
\langle \zeta ({\bf k}_1) \zeta ({\bf k}_2) \zeta ({\bf k}_3) \rangle_{\cal D} = 0
\label{zzzDfinal}
\eeq
identically in the equilateral limit.

The new nonvanishing terms in Eqs. (\ref{zzzAfinal}) and (\ref{zzzCfinal}) remain considerably smaller than the $f_{NL}$ term of Eq. (\ref{zzzfNL}) for the family of models of interest. The term stemming from ${\cal A}^{IJK}$ in Eq. (\ref{zzzAfinal}) is proportional to $(N_{, A} N^{, A} )$, whereas the $f_{NL}$ term is multiplied to the square of that term. For models of interest here, in which the potential includes ridges, the gradient term is significant. For each of the three trajectories of Fig. 2, for example, $(N_{, A} N^{, A}) = {\cal O} (10^3)$ across the full range $N_* = 60$ to $N_* = 40$. The gradient increases as the ratio of $\xi_\chi / \xi_\phi$ increases, and hence the $f_{NL}$ term will dominate the term coming from ${\cal A}^{IJK}$ whenever $\vert f_{NL} \vert > 10^{-3}$. 

For the term involving ${\cal R}^{IABJ; K}$ in Eq. (\ref{zzzCfinal}), we may take advantage of the fact that for two-field models the Riemann tensor for the field space may be written
\beq
{\cal R}_{ABCD} = {\cal K} (\phi^I) \left[ {\cal G}_{AC} {\cal G}_{BD} - {\cal G}_{AD} {\cal G}_{BC} \right] ,
\label{Riemann2d}
\eeq
where ${\cal K} (\phi^I )$ is the Gaussian curvature. In two dimensions, ${\cal K} (\phi^I ) = { 1 \over 2 } {\cal R} (\phi^I)$, where ${\cal R}$ is the Ricci scalar. Since ${\cal D}_K {\cal G}_{AB} = {\cal G}_{AB; K} = 0$ and ${\cal K} (\phi^I)$ is a scalar in the field space, the covariant derivative of the Riemann tensor is simply proportional to the ordinary (partial) derivative of the Gaussian curvature, ${\cal K}$. In particular, we find
\beq
\hat{\sigma}_A \hat{\sigma}_B N_{, I} N_{, J} N_{, K} {\cal R}^{IABJ; K} = - \left( \hat{s}^{IJ} N_{, I} N_{, J} \right) \left( N_{, K} {\cal K}^{, K} \right) ,
\label{Rgradient}
\eeq
where $\hat{s}^{IJ} \equiv {\cal G}^{IJ} - \hat{\sigma}^I \hat{\sigma}^J$ is the projection operator for directions orthogonal to the adiabatic direction. We calculate ${\cal K}$ in Eq. (\ref{Ricci2d}). At early times, as the system undergoes slow-roll inflation, we have $\xi_\phi \phi^2 + \xi_\chi \chi^2 \gg M_{\rm pl}^2$. For the trajectories as in Fig. 2, moreover, the system evolves along a ridge such that $\xi_\phi \phi^2 \gg \xi_\chi \chi^2$. In that case, we find
\beq
{\cal K} \simeq \frac{1}{108 \xi_\phi^2 M_{\rm pl}^2 } \left[ 1 + 6 (\xi_\phi + \xi_\chi) + 36 \xi_\phi (\xi_\chi - \xi_\phi ) \right] \sim \phi^0 , \> \chi^0 ,
\label{gradK}
\eeq
and hence ${\cal K}^{, I} \sim 0$. Thus, in addition to being suppressed by the slow-roll factor, $\epsilon$, the contribution to the primordial bispectrum from the ${\cal R}^{IABJ; K}$ term is negligible in typical scenarios of interest, because of the weak variation of the Gaussian curvature of the field-space manifold around the times $N_* = 60$ to $N_* = 40$ efolds before the end of inflation. This matches the behavior shown in Fig. \ref{Rphichi}: the field-space manifold is nearly flat until one reaches the vicinity of $\phi, \chi \sim 0$, near the end of inflation.

Though these results were derived in the equilateral limit, for which $k_1 =  k_2 =  k_3 = k_*$, we expect the same general pattern to apply more generally, for example, to the squeezed local configuration in which $k_1 \simeq  k_2  = k_*$ and $k_3  \simeq 0$. As one departs from the equilateral limit the exact cancellations of Eqs. (\ref{zzzBfinal}) and (\ref{zzzDfinal}) no longer hold, though each of the components of the field-space Riemann tensor and its gradients remains small between $N_* = 60$ to $N_* = 40$ efolds before the end of inflation for models of the class we have been studying here. Meanwhile, the $k$-dependent functions, $f_{\cal I} ({\bf k}_i)$ in Eq. (\ref{ABCD}), remain of comparable magnitude to the $k$-dependent contribution in Eq. (\ref{zzzfNL}) \cite{SeeryBispectrum} --- each contributes as $\left[ {\cal O} (1) - {\cal O} (10) \right] \times k^3$ --- while the coefficients of the additional terms arising from ${\cal A}^{IJK}$, ${\cal B}^{IJK}$, ${\cal C}^{IJK}$, and ${\cal D}^{IJK}$ are further suppressed by factors of $\epsilon$. For models of the class we have been studying here, we therefore expect the (covariant version of the) usual $f_{NL}$ term to dominate the primordial bispectrum. Moreover, given the weak dependence of the Gaussian curvature ${\cal K} (\phi^I)$ with $\phi^I$ between $N_* = 60$ and $N_* = 40$, we do not expect any significant contributions to the running of $f_{NL}$ with scale to come from the curvature of the field-space manifold, given the analysis in \cite{ByrnesGong}.

We calculate the magnitude of $f_{NL}$ numerically, following the definition in Eq. (\ref{fNL}). The discrete derivative of $N$ along the $\phi$ direction is constructed as
\beq
N_{, \phi} = \frac{ N (\phi + \Delta \phi , \chi) - N (\phi - \Delta \phi, \chi ) }{2 \Delta \phi } ,
\label{Nderiv}
\eeq
where $N (\phi, \chi)$ is the number of efolds between $t_*$ and $t_{\rm end}$, where $t_{\rm end}$ is determined by the physical criterion that $\ddot{a} = 0$ (equivalent to $\epsilon = 1$). For each quantity, such as $N (\phi + \Delta \phi, \chi)$, we re-solve the exact background equations of motion numerically and measure how the small variation in field values at $t_*$ affects the number of efolds of inflation between $t_*$ and the time at which $\ddot{a} = 0$. The discrete derivatives along the other field directions and the second derivatives are constructed in a corresponding manner. Covariant derivatives are calculated using the discrete derivatives defined here and the field-space Christoffel symbols evaluated at background order. For the trajectories of interest, the fields violate slow-roll late in their evolution (after they have fallen off the ridge of the potential), but they remain slowly rolling around the time $t_*$; if they did not, as we saw in Section IV, then the predictions for the spectral index, $n_s (t_*)$ would no longer match observations. We therefore do not consider separate variations of the field velocities at the time $t_*$, since in the vicinity of $t_*$ they are related to the field values. Because the second derivatives of $N$ are very sensitive to the step sizes $\Delta \phi$ and $\Delta \chi$, we work with 32-digit accuracy, for which our numerical results converge for finite step-sizes in the range $\Delta \phi , \Delta \chi = \{ 10^{-6}, 10^{-5} \}.$

For the three trajectories of Fig. \ref{trajectories123}, we find the middle case, trajectory 2, yields a value of $f_{NL}$ of particular interest: $\vert f_{NL} \vert = 43.3$ for fiducial scales $k_*$ that first crossed the Hubble radius $N_* = 60$ efolds before the end of inflation. Note the strong sensitivity of $f_{NL}$ to the fields' initial conditions: varying the initial value of $\chi (\tau_0)$ by just $\vert \Delta \chi (\tau_0 ) \vert = 10^{-3}$ changes the fields' evolution substantially --- either causing the fields to roll off the hill too early (trajectory 1) or not to turn substantially in field space at all (trajectory 3) --- both of which lead to negligible values for $f_{NL}$. See Fig. \ref{fNL1}.
\begin{figure}
\centering
\includegraphics[width=4.5in]{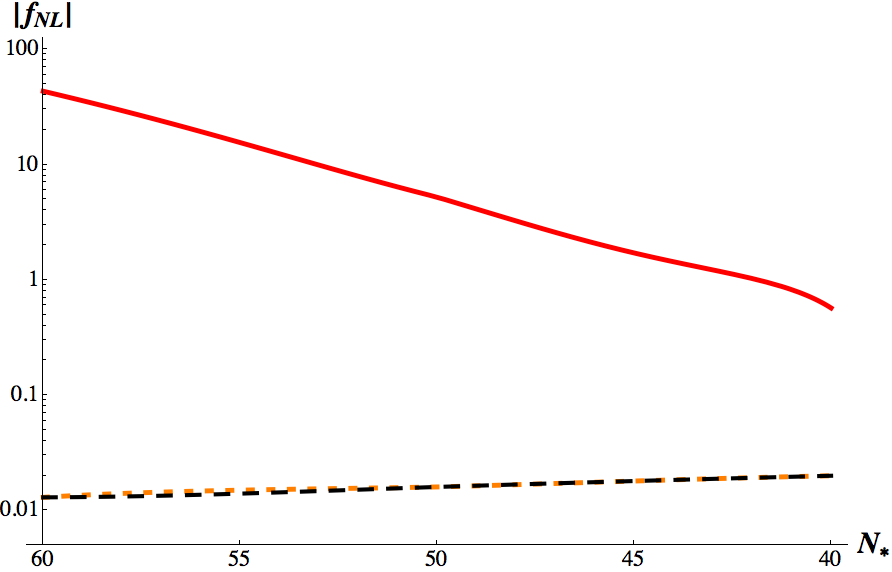}
\caption{\small \baselineskip 14pt The non-Gaussianity parameter, $\vert f_{NL} \vert$, for the three trajectories of Fig. \ref{trajectories123}: trajectory 1 (orange dotted line); trajectory 2 (solid red line); and trajectory 3 (black dashed line). Changing the fields' initial conditions by a small amount leads to dramatic changes in the magnitude of the primordial bispectrum.}
\label{fNL1}
\end{figure}

\section{Conclusions}

We have demonstrated that multifield models with nonminimal couplings generically produce the conditions required to generate primordial bispectra of observable magnitudes. Such models satisfy at least three of the four criteria identified in previous reviews of primordial non-Gaussianities \cite{WhitePaper,XingangReview}, namely, the presence of multiple fields with noncanonical kinetic terms whose dynamics temporarily violate slow-roll evolution.

Two distinct features are relevant in this class of models: the conformal stretching of the effective potential in the Einstein frame, which introduces nontrivial curvature distinct from features in the Jordan-frame potential; and nontrivial curvature of the induced manifold for the field space in the Einstein frame. So long as the nonminimal couplings are not precisely equal to each other, the Einstein-frame potential will include bumps or ridges that will tend to cause neighboring trajectories of the fields to diverge over the course of inflation. Such features of the potential are generic to this class of models, and hence are strongly motivated by fundamental physics. 

We have found that the curvature of the potential dominates the effects of interest at early and intermediate stages of inflation, whereas the curvature of the field-space manifold becomes important near the end of inflation (and hence during preheating). The generic nature of the ridges in the Einstein-frame potential removes one of the kinds of fine-tuning that have been emphasized in recent studies of non-Gaussianities in multifield models, namely, the need to introduce potentials of particular shapes \cite{ByrnesReview,TanakaMultifield,SeeryFocusing,PTGeometric}. (We are presently performing an extensive sweep of parameter space to investigate how $f_{NL}$ behaves as one varies the couplings $\xi_I$, $\lambda_I$, and $m_I$. This will help determine regions of parameter space consistent with current observations.) On the other hand, much as in \cite{ByrnesReview,TanakaMultifield,SeeryFocusing,PTGeometric}, we find a strong sensitivity of the magnitude of the bispectrum to the fields' initial conditions. Thus the production during inflation of bispectra with magnitude $\vert f_{NL} \vert \sim {\cal O} (50)$ requires fine-tuning of initial conditions such that the fields begin at or near the top of a ridge in the potential.

A subtle question that deserves further study is whether the formalism and results derived in this paper show any dependence on frame. Although we have developed a formalism that is gauge-invariant with respect to spacetime gauge transformations, and covariant with respect to the curvature of the field-space manifold, we have applied the formalism only within the Einstein frame. The authors of \cite{Sasaki} recently demonstrated that gauge-invariant quantities such as the curvature perturbation, $\zeta$, can behave differently in the Jordan and Einstein frames for multifield models with nonminimal couplings. The question of possible frame-dependence of the analysis presented here remains under study. Whether quantities such as $f_{NL}$ show significant evolution during reheating for this family of models, as has been emphasized for related models \cite{SeeryFocusing,LeungByrnes}, likewise remains a subject of further research.

\section*{Appendix A: Field-Space Metric and Related Quantities}

Given $f (\phi^I)$ in Eq. (\ref{f2field}) for a two-field model, the field-space metric in the Einstein frame, Eq. (\ref{GIJ}), takes the form
\beq
\begin{split}
{\cal G}_{\phi \phi} &= \left( \frac{M_{\rm pl}^2}{2f} \right) \left[ 1 + \frac{3 \xi_\phi^2 \phi^2}{f} \right] , \\
{\cal G}_{\phi \chi} = {\cal G}_{\chi \phi} &= \left( \frac{M_{\rm pl}^2}{2f} \right) \left[ \frac{3 \xi_\phi \xi_\chi \phi \chi}{f} \right] , \\
{\cal G}_{\chi \chi} &= \left( \frac{M_{\rm pl}^2}{2f} \right) \left[ 1 + \frac{3 \xi_\chi^2 \chi^2}{f} \right] .
\end{split}
\label{Gphiphi}
\eeq
The components of the inverse metric are
\beq
\begin{split}
{\cal G}^{\phi \phi} &= \left( \frac{ 2f}{M_{\rm pl}^2} \right) \left[ \frac{2f + 6 \xi_\chi^2 \chi^2}{C} \right] , \\
{\cal G}^{\phi \chi} = {\cal G}^{\chi \phi} &= - \left( \frac{ 2f}{M_{\rm pl}^2} \right) \left[ \frac{6 \xi_\phi \xi_\chi \phi \chi }{C} \right] , \\
{\cal G}^{\chi\chi} &= \left( \frac{ 2f}{M_{\rm pl}^2} \right)  \left[ \frac{2f + 6 \xi_\phi^2 \phi^2}{C} \right] ,
\end{split}
\label{inverseG}
\eeq
where we have defined the convenient combination
\beq
\begin{split}
C (\phi, \chi) &\equiv M_{\rm pl}^2 + \xi_\phi (1 + 6 \xi_\phi) \phi^2 + \xi_\chi (1 + 6 \xi_\chi) \chi^2 \\
&= 2f + 6 \xi_\phi^2 \phi^2 + 6 \xi_\chi^2 \chi^2 .
\label{C}
\end{split}
\eeq

The Christoffel symbols for our field space take the form
\beq
\begin{split}
\Gamma^\phi_{\>\> \phi \phi} &= \frac{\xi_\phi (1 + 6 \xi_\phi ) \phi}{C} - \frac{\xi_\phi \phi}{f} , \\
\Gamma^\phi_{\>\> \chi \phi} = \Gamma^\phi_{\>\> \phi \chi} &= - \frac{\xi_\chi \chi}{2f} , \\
\Gamma^\phi_{\>\> \chi \chi} &= \frac{\xi_\phi (1 + 6 \xi_\chi) \phi}{C} , \\
\Gamma^\chi_{\>\> \phi \phi} &= \frac{\xi_\chi (1 + 6 \xi_\phi) \chi}{C} , \\
\Gamma^\chi_{\>\> \phi \chi} = \Gamma^\chi_{\>\> \chi \phi} &= - \frac{\xi_\phi \phi}{2f} , \\
\Gamma^\chi_{\>\> \chi \chi} &= \frac{\xi_\chi (1 + 6 \xi_\chi ) \chi}{C} - \frac{\xi_\chi \chi}{f}
\end{split}
\label{Gammas}
\eeq

For two-dimensional manifolds we may always write the Riemann tensor in the form
\beq
{\cal R}_{ABCD} = {\cal K} (\phi^I) \left[ {\cal G}_{AC} {\cal G}_{BD} - {\cal G}_{AD} {\cal G}_{BC} \right] ,
\label{Riemann2d}
\eeq
where ${\cal K} (\phi^I)$ is the Gaussian curvature. In two dimensions, ${\cal K} (\phi^I ) = { 1 \over 2} {\cal R} (\phi^I)$, where ${\cal R} (\phi^I)$ is the Ricci scalar. Given the field-space metric of Eq. (\ref{Gphiphi}), we find
\beq
{\cal R} (\phi^I ) = 2 {\cal K} (\phi^I) = \frac{2}{3 M_{\rm pl}^2 C^2} \left[ (1 + 6 \xi_\phi) (1 + 6 \xi_\chi) (4 f^2 ) - C^2 \right] .
\label{Ricci2d}
\eeq

\acknowledgements{It is a great pleasure to thank Mustafa Amin, Bruce Bassett, Rhys Borchert, Xingang Chen, Joseph Elliston, Alan Guth, Carter Huffman, Francisco Pe\~{n}a, Katelin Schutz, and David Seery for helpful discussions. This work was supported in part by the U.S. Department of Energy (DoE) under contract No. DE-FG02-05ER41360. EAM was also supported in part by MIT's Undergraduate Research Opportunities Program (UROP).}


\begin{thebibliography}{999}
\bibitem{Footnote} The leading behavior of Eq. (\ref{Vchichi}) in the limit $\xi_\phi \phi^2 \gg M_{\rm pl}^2$ is unchanged if we calculate the curvature of the potential using the covariant derivative for the field-space, $({\cal D}_\chi^2 V )_{\vert \chi = 0}$, where ${\cal D}_I$ is defined in Eq. (\ref{covderiv1}).
\bibitem{Footnote2} Because the field fluctuations at time $t_1$ live in a different tangent space to the field-space manifold than do the fluctuations at time $t_2$, the authors of \cite{SeeryBispectrum} introduce propagators and related transport functions to translate between quantities in the two distinct tangent spaces, building on previous work in \cite{SeeryTransfer}. Since the two tangent spaces coincide in the limit $t_2 \rightarrow t_1$, the propagators reduce to the identity matrix in that limit. What matters in our discussion is the symmetry structure of the various terms in Eq. (\ref{QQQ1}), which should hold at any time, hence we evaluate the three-point function at time $t_*$.
\bibitem{Footnote3} We are especially indebted to Joseph Elliston for helpful discussions on this point.
\end{thebibliography}
\end{document}